\documentclass[a4paper,11pt,dvips]{article}
cm \voffset = -2truecm

\usepackage{graphicx}
\usepackage{amsmath}
\usepackage{amssymb}
\usepackage{subcaption}
\usepackage{latexsym}
\usepackage{color}
\usepackage{float}
\usepackage{cite}
\usepackage{makeidx}
\usepackage{cancel}
\usepackage[dvipdfm,colorlinks]{hyperref}
\restylefloat{figure}

\newcommand{\bra}{\begin{array}}
\newcommand{\era}{\end{array}}
\newcommand{\beq}{\begin{equation}}
\newcommand{\eeq}{\end{equation}}
\newcommand{\bqr}{\begin{eqnarray}}
\newcommand{\eqr}{\end{eqnarray}}

\def\BC{\bb C}
\def\_\BC{\bbi C}

\def\( {\left(}
   \def\) {\right)}
\def\[ {\left[}
\def\] {\right]}
\def\no2 {{\textstyle{n\over 2}}}

\newcommand{\si}{\sigma}

\newcommand{\te}{\theta}

\begin{document}
\begin{titlepage}
\setcounter{page}{1}
\renewcommand{\thefootnote}{\fnsymbol{footnote}}

\begin{flushright}
%ucd-tpg **-**-****\\
\end{flushright}

\vspace{5mm}
\begin{center}

{\Large \bf Factorization of Dirac Equation and Graphene Quantum Dot}

\vspace{5mm}

{\bf Youness Zahidi}$^{a}$,
 {\bf Ahmed Jellal\footnote{\sf ajellal@ictp.it --
a.jellal@ucd.ac.ma}}$^{a,b}$,  {\bf Hocine Bahlouli}$^{b,c}$
and {\bf Mohammed El Bouziani}$^{a}$

\vspace{5mm}

{$^{a}$\em Theoretical Physics Group,  %Department of Physics,
Faculty of Sciences, Choua\"ib Doukkali University},\\
%{\em PO Box 20,
{\em 24000 El Jadida, Morocco}

{$^{b}$\em Saudi Center for Theoretical Physics, Dhahran, Saudi
Arabia}

{$^{c}$\em Physics Department, King Fahd University of Petroleum $\&$ Minerals, \\
Dhahran 31261, Saudi Arabia}

\vspace{3cm}

\begin{abstract}

We consider a quantum dot described by a cylindrically symmetric
2D Dirac equation. The potentials representing the quantum dot are
taken to be of different types of potential configuration, scalar,
vector and pseudo-scalar to enable us to enrich our study. Using
various potential configurations, we found that in the presence of
a mass term an electrostatically confined quantum dot can accommodate
true bound states, which is in agreement with previous work. The
differential cross section associated with one specific potential
configuration has been computed and discussed as function of the
various potential parameters.

\end{abstract}
\end{center}
\vspace{5cm}


\end{titlepage}

%%%%%%%%%%%%%%%%%%%%%%%%%%%%%%%%%%%%%%%%%%%%%%%%%%%
\section{Introduction}
%%%%%%%%%%%%%%%%%%%%%%%%%%%%%%%%%%%%%%%%%%%%%%%

 Recent technological advances in nanofabrication have
created a great deal of interest in the study of low dimensional
quantum systems such as quantum wells, quantum wires, and quantum
dots. In particular, there has been considerable amount of work in
recent years on confined semiconductor structures, which finds
applications in electronic and optoelectronic devices. The two
dimensional character of the system allows for electron confinement in one
spatial direction for a specific potential configuration.
Graphene \cite{graph, graph2} %, a single sheet of carbon  atoms
%arranged in a two-dimensional (2D) honeycomb crystal structure,
has became one of the most important subjects in condensed matter research in the
last few years. This is because of its exotic physical properties and
the apparent similarity of its mathematical model to the one
describing relativistic fermions in 2-dimensions (2D). Graphene is a fascinating subject
because its low energy quasiparticles are governed by a (2+1)-dimensional
Dirac equation with the Fermi velocity $v_F$. These
unique and amazing properties make graphene one of the most
promising materials for future nanoelectronics devices
\cite{Geim}.
Graphene quantum dots (artificial atoms) \cite{Chak, Silves,
Matul} have ignited intense research activities related to quantum
information storage and processing using spin information of the confined
electrons. Various methods were used to make quantum dots (QD), one of the
most widely used techniques uses electrostatic gates \cite{Silves}.

On the other hand, the main features of the conductivity of doped single layer graphene were
analyzed and models for different scattering mechanisms were presented by Guinea~\cite{Guinea}. Many possible
dependencies of the cross section on the Fermi wavelength were identified, depending
on the type of scattering mechanism. Defects with internal structure, such
as ripples, showed non monotonous dependencies, with maxima when the Fermi wavelength
is comparable to the typical length scale of the defect.
Furthermore, the electronic states of an electrostatically confined cylindrical graphene quantum dot and the electric transport
through this device were studied theoretically within the continuum Dirac-equation approximation and compared
with numerical results obtained from a tight-binding lattice description by Pal {\it et al.} \cite{Pal}. A spectral gap, which may originate
from strain effects, additional adsorbed atoms, or substrate-induced sublattice-symmetry breaking, allowed for
bound and scattering states. As long as the diameter of the dot is much larger than the lattice constant, the results
of the continuum and the lattice model are in very good agreement. The influence of
dot-potential step, on-site disorder along the sample edges, uncorrelated short-range disorder potentials in
the bulk and of random magnetic fluxes that mimic ripple disorder, were investigated. It was concluded that
the quantum dots spectral and transport properties depend crucially on the specific type of disorder and in general,
the peaks in the density of bound states are broadened but remain sharp only in the case of edge disorder.

Very recently, we have presented a systematic approach for the separation of variables for the two-dimensional Dirac equation
in polar coordinates \cite{paper1}.
The three vector potential, which couple to the Dirac spinor via minimal coupling, along with the scalar potential were chosen
to have angular dependence which emanate the Dirac equation to complete separation of variables. Exact solutions were obtained
for a class of solvable potentials along with their relativistic spinor wave functions. Particular attention was paid to the situation
where the potentials were confined to a quantum dot region and were of scalar, vector and pseudo-scalar type. The study of a single charged
impurity embedded in a 2D Dirac equation in the presence of a uniform magnetic field was treated as a particular case of our general study.

In this paper we use our recently developed formalism for the 2D Dirac equation \cite{paper1}
and apply our results to graphene based on the recent results reported in \cite{Guinea, Pal}. In particular, we study the energy spectrum
of graphene QD in a presence of an electrostatic confining potential.
One of our purpose %of this paper
is to study the elastic scattering theory
through  radially-symmetric potentials and evaluate the transport cross section
which is very valuable for the study of the transport properties of graphene \cite{scat}.
In order to probe the transport properties of the QD we add an environment
to the isolated QD so that the exponentially decaying bound states are still
finite when reaching the outer region. We explicitly investigate the
electronic transport through a circular electrostatic potential in
the presence of a constant mass term. We first obtain the
asymptotic form of the 2D wave function, we write them in terms of normalized
spinor plane and cylindrical waves. Using the definition of the
scattering matrix \cite{Guinea, Kats} we calculate the differential
cross section and then deduce the transport cross section, and
finally we draw our conclusions.

The paper is organized as follows. Section 2 summarizes our separation of variables approach for 2D
Dirac equation which was used in our previous work. In section 3, we give the solutions of the energy spectrum  of the Dirac
equation for two potential configurations. In section 4, we concentrate on graphene and assume the presence of
a constant mass term that can be induced by different
experimental methods \cite{Zhou}. We include a radially symmetric
potential defining the QD and obtain the solution of the associated
Dirac equation in various regions. The resulting energies of the bound states
of the isolated QD are in good agreement with previous calculations of bound state
energy \cite{Pal, Nilsson}. We note that various methods have been used to analyze
bound states in graphene \cite{martino, Pal, Nilsson}.
In section 5, we study the transport properties for graphene quantum dot by
evaluating and studying the corresponding cross sections.
Explicit investigation of scattering processes in graphene will be performed for a specific
potential configuration in section 6. We finally conclude our work in last section.

%%%%%%%%%%%%%%%%%%%%%%%%%%%%%%%%%%%%%%%%%%%%%%%%%%%
\section{Theoretical model}
%%%%%%%%%%%%%%%%%%%%%%%%%%%%%%%%%%%%%%%%%%%%%%%

In this section we start by summarizing the main steps involved in the separation of variables approach we used
for 2D Dirac equation \cite{paper1}. Consider the 2D Dirac equation with an electromagnetic
interaction through minimal coupling for a spin 1/2 particle of
mass \textit{m} and charge \textit{e} in units  $({\rm
\hslash }=c=1)$
\begin{equation} \label{GEQ111}
\left[\gamma ^{\mu } \left(i\partial _{\mu } -eA_{\mu }
\right)-\left(m+S\right) \right] \psi =0
\end{equation}
where $\gamma ^{\mu } \partial _{\mu } =\gamma ^{0} \partial _{0}
+\vec{\gamma }\cdot \vec{\nabla }$, $S$ is the pseudo-vector
coupling and $A_{\mu } =(A_{0} ,\vec{A})$ with $A_{0} $ is related
to the electrostatic potential $\vec{E}=-\vec{\nabla }A_{0}
-\frac{\partial \vec{A}}{\partial t} $ and $\vec{A}$ is related to the
magnetic field $\vec{B}=\vec{\nabla }\times \vec{A}$. The Dirac
matrices $\gamma^{\mu}$ satisfy the algebra
\begin{equation}
\left[  \gamma^{\mu},\gamma^{\nu}\right]
=-2i\sigma^{\mu\nu},\qquad \left\{
\gamma^{\mu},\gamma^{\nu}\right\} =2\eta^{\mu\nu}
\end{equation}
with $\eta^{\mu\nu}=\mbox{diag}\left(  1,-1,-1\right)$ and
$\mu,\nu=0,1,2$. In (2+1)-dimensions we select the following
representation $\gamma ^{0} =i\sigma _{3}$, $\vec{\gamma}
=i\vec{\sigma}$, where $\{\sigma_i\}_{i=1}^{3}$ are the $2\times2$
Pauli matrices
\begin{equation} \label{GEQ112}
\begin{array}{l} {\quad \quad \sigma _{1} =\left(\begin{array}{cc} {0} & {1} \\ {1} & {0} \end{array}\right),
\qquad  \sigma _{2} =\left(\begin{array}{cc} {0} & {-i}
\\ {i} & {0} \end{array}\right),\qquad \sigma _{3}
=\left(\begin{array}{cc} {1} & {0} \\ {0} & {-1}
\end{array}\right)}
\end{array}
\end{equation}
and then \eqref{GEQ111} can be written as follows
\begin{equation} \label{GEQ114}
i\gamma ^{0} \left(\frac{\partial }{\partial t} \Psi
\right)+i\vec{\gamma }\cdot \vec{\nabla }\Psi -{\rm e}\vec{\gamma
} \cdot \vec{A}\Psi -\left(m+S\right)\Psi -{\rm e}\gamma ^{0}
A_{0} =0.
\end{equation}
 Multiplying \eqref{GEQ114} by $\gamma ^{0}$ and using the notation
 $\vec{\alpha }= \gamma ^{0} \vec{\gamma }$, $\beta =\gamma ^{0}$
 %we obtain a Schrodinger-like equation
to obtain
\begin{equation} \label{GEQ115}
i\frac{\partial }{\partial t} \psi =\left[-i\vec{\alpha } \cdot
\vec{\nabla }+eA_{0} +e\vec{\alpha } \cdot
\vec{A}+\left(m+S\right)\beta \right] \psi =H \Psi.  \end{equation}
In the forthcoming analysis we study different potential
configurations in order to solve explicitly the above equation. For
time-independent potentials, the two components spinor
wave function can be written as follows $\Psi (t,r,\theta
)=e^{-i\varepsilon t} \Psi (r,\theta )$ so that our previous
equation becomes
\begin{equation} \label{GEQ118}
 (H-\varepsilon )\Psi (r,\theta )=0.
 \end{equation}
Knowing that in polar coordinates $\vec{\nabla
}=\hat{r}\frac{\partial }{\partial r} +\hat{\theta }\frac{1}{r}
\frac{\partial }{\partial \theta } $ and $\vec{\alpha }=i\sigma
_{3} \vec{\sigma }$ so that our Hamiltonian is now
\begin{equation} \label{GEQ120}
H=eA_{0} +\left(m+S\right)\sigma _{3} +\sigma _{3} \sigma _{r}
\partial _{r} +ie\sigma _{3} \sigma _{r} A_{r}
+\sigma _{3} \sigma _{\theta } \frac{1}{r} \partial _{\theta }
+ie\sigma _{3} \sigma _{\theta } A_{\theta }.
\end{equation}
To proceed further we consider, along the line of our previous
paper \cite{jellal}, a unitary transformation $\Lambda (r,\theta
)$ that transform $(\si_r,\si_\te)$ into $(\si_1,\si_2)$ and vice versa.
Thus we require that
\begin{equation} \label{GEQ123}
\Lambda \sigma _{r} \Lambda ^{-1} =\sigma _{1} ,\qquad  \Lambda
\sigma _{\theta } \Lambda ^{-1} =\sigma _{2}
\end{equation}
 %and vice versa.
which then turns out to have the following explicit form
\begin{equation} \label{GEQ124}
\Lambda (r,\theta )=\lambda (r,\theta )e^{\frac{i}{2} \sigma _{3}
\theta }
\end{equation}
where $\lambda (r,\theta )$ is a $1 \times 1$ real function and the
exponential is a $2 \times 2$ unitary matrix. Then we can define
the new Hamiltonian in matrix form as
\begin{equation} \label{GEQ131}
{\rm {\mathcal H}}=
\begin{pmatrix}
  {m+S+eA_{0} } & {\partial _{r} -
\frac{\lambda _{r} }{\lambda } +\frac{1}{2r} +ieA_{r} -\frac{i}{r}
(\partial _{\theta } -\frac{\lambda _{\theta } }{\lambda }
)+eA_{\theta } } \\ {-\partial _{r} +\frac{\lambda _{r} }{\lambda
} -\frac{1}{2r} -ieA_{r} -\frac{i}{r} (\partial _{\theta }
-\frac{\lambda _{\theta } }{\lambda } )+eA_{\theta } } &
{-m-S+eA_{0} }
  \end{pmatrix}.
  \eeq
One can show that  the hermiticity of ${\rm {\mathcal H}}$
requires  $\lambda =\sqrt{r} $ and reduces the Hamiltonian to
\begin{equation} \label{GEQ135}
{\rm {\mathcal H}}=\left(\begin{array}{cc} {m+S+eA_{0} } &
{\partial _{r} +ieA_{r} -\frac{i}{r} \partial _{\theta }
+eA_{\theta } } \\ {-\partial _{r} -ieA_{r} -\frac{i}{r} \partial
_{\theta } +eA_{\theta } } & {-m-S+eA_{0} } \end{array}\right)
\end{equation}
and we obtain the $(2+1)$-dimensional Dirac equation
$\left({\rm {\mathcal H}}-\varepsilon \right)\chi =0$ or equivalently
\begin{equation} \label{GEQ136}
\left(\begin{array}{cc} {m+S+eA_{0} -\varepsilon } & {\partial
_{r} +ieA_{r} -\frac{i}{r} \partial _{\theta } +eA_{\theta } } \\
{-\partial _{r} -ieA_{r} -\frac{i}{r} \partial _{\theta }
+eA_{\theta } } & {-m-S+eA_{0} -\varepsilon }
\end{array}\right)\left(\begin{array}{c} {\chi _{+} (r,\theta )}
\\ {\chi _{-} (r,\theta )} \end{array}\right)=0
\end{equation}
where the transformed spinor wavefunction,
$\chi(r,\theta)=(\chi_{+}(r,\theta),\chi_{-}(r,\theta))^t$ and the
superscript $t$ stands for transpose of the spinor, is given by
\begin{equation}
\chi(r,\theta)=\sqrt{r} e^{\frac{i}{2}\sigma_3 \theta}\Psi
(r,\theta ).
\end{equation}
These will be used to explicitly determine the solutions of the energy spectrum,
which will serve to deal with different issues.

%%%%%%%%%%%%%%%%%%%%%%%%%%%%%%%%%%%%%%%%%%%%%%%%%%%
\section{Potential configurations and solutions} %Wavefunctions}
%%%%%%%%%%%%%%%%%%%%%%%%%%%%%%%%%%%%%%%%%%%%%%%

In order to determine the energy spectrum, we use
the potential configurations which were used in our work
\cite{paper1} that ensure separation of variables.
Now we can write the spinor wave function as
$\chi_{\pm}(r,\theta)=\Phi(r)_{\pm}F(\theta)_{\pm}$ where the subscripts
stand for upper and lower spinor components. We consider the first potential
configuration defined by
\beq
A_{0}(\vec{r})=V(r), \qquad  A_{r}(\vec{r})=R(r), \qquad
A_{\theta}=W(r), \qquad S=S(r). \label{first-pc}
\eeq
We note that the radial part of the vector potential $A_r$ can be gauged away
in the above situations and hence will not be included in our
future equations, \eqref{GEQ136} becomes
\beq\label{eq15} \left[
  \left(\begin{array}{cc}
    {m+S+eV -\varepsilon} & {\partial _{r} + eW} \\
    {-\partial _{r} + eW} & {-m-S+eV -\varepsilon} \\
  \end{array}\right)
  +\frac{1}{r}\left(
                \begin{array}{cc}
                  {0} & {-i\partial _{\theta }} \\
                  {-i\partial _{\theta }} & {0} \\
                \end{array}
              \right)
\right] \left(
  \begin{array}{c}
    {\Phi_{+}F_{+}} \\
    {\Phi_{-}F_{-}} \\
  \end{array}
\right)=0.
\eeq
The angular component satisfies
$\partial_{\theta}F(\theta)=i\varepsilon_{\theta}F(\theta)$,
giving the solution
\beq F(\theta) =e^{i\varepsilon_{\theta}\theta}\eeq
where the parameter $\varepsilon_{\theta}$ will be defined later by the boundary conditions. As a result,
the following differential equation for each spinor component is obtained
 \beq\label{eq33}
 \left[\frac{d^{2}}{dx^{2}}-\frac{\mu_{\pm}^2-\frac{1}{4}}{x^{2}}+\frac{\nu}{x}-\frac{1}{4}\right]\Phi _{\pm} =0
\eeq
where we have defined the variable $x=2\gamma r$ and the three quantities $\nu=-\frac{eW
\varepsilon_{\theta}}{\gamma}$,
$\mu_{\pm}^2=\left(\varepsilon_{\theta}\mp \frac{1}{2}\right)^2$,
$\gamma^{2}=\left(m+S\right)^{2}+e^2W^2-\left(\varepsilon-eV\right)^{2}$.

The radial equation \eqref{eq33} is solved in term of the
Whittaker hypergeometric functions $M_{\nu,\mu_{\pm}}(2\gamma r)$ and
$W_{\nu,\mu_{\pm}}(2\gamma r)$. The general solution takes the form \beq
\label{sol1} \Phi _{\pm} (r) = A_\pm M_{\nu,\mu_{\pm}}(2\gamma r) +
B_{\pm} W_{\nu,\mu_{\pm}}(2\gamma r) \eeq
where $M_{\nu,\mu_{\pm}}(2\gamma r)$ and $W_{\nu,\mu_{\pm}}(2\gamma r)$ are given in terms of confluent
hypergeometric functions \cite{Abramowitz} \bqr
&& M_{\nu,\mu_{\pm}}(2\gamma r)= e^{-\gamma r} (2\gamma r)^{{\mu_{\pm}} +1/2} {}_1F_1(1/2 +\mu_{\pm} - \nu,1+2\mu_{\pm},2\gamma r)\\
&& W_{\nu,\mu_{\pm}}(2\gamma r)= e^{-\gamma r} (2\gamma r)^{\mu_{\pm} +1/2}
U(1/2 +\mu_{\pm} - \nu,1+2\mu_{\pm},2\gamma r). \eqr
The general solution to our original problem can be written as follows
 \beq \Psi (r,\theta
)=\frac{1}{\sqrt{r}}e^{i(\varepsilon_{\theta}- \frac{1}{2} \sigma
_{3})\theta }\Phi(r). \eeq
On the other hand, the boundary condition on the total wave function
$\psi(r,\theta)=\psi(r,\theta+2\pi)$ requires that $e^{i(2\pi
\varepsilon_{\theta}-\sigma_3 \pi)}=1$ which gives the following
quantization rule for the parameter $\varepsilon_{\theta}$
\beq \varepsilon_{\theta}=\frac{k}{2}, \qquad
k=\pm1, \pm3, \pm5,\cdots.
\eeq
Hence the most general solution of our problem reads
 \beq \Psi_{\pm} (r, \theta)=\sum_{k,\pm}
\frac{1}{\sqrt{r}}e^{\frac{i}{2}(k - \sigma _{3}) \theta } \left[A_{\pm}
M_{\nu,\mu_{\pm}}(2\gamma r) +B_{\pm} W_{\nu,\mu_{\pm}}(2\gamma r) \right] \eeq
which are also eigenfunctions of the total angular momentum
\beq
J_z =L_z + \frac{1}{2}\sigma _{3}=-i
\partial_{\te}+ \frac{1}{2}\sigma _{3}.
\eeq

A second potential configuration can be considered in our present work. It
is defined by the  potential parameters
%Now, we study the case where the  potential configuration has the
%following configuration
\beq A_{0}(\vec{r})=V(r), \qquad
A_{r}(\vec{r})=R(r), \qquad A_{\theta}=\frac{W(\theta)}{r}, \qquad
S=S(r).\label{eq18}
\eeq
In this case \eqref{GEQ136} can be written as
\beq \label{eq19} \left[
   \begin{pmatrix}
    {m+S+eV -\varepsilon} & {\partial _{r} +ieR} \\
    {-\partial _{r} -ieR} & {-m-S+eV -\varepsilon} \\
  \end{pmatrix}
  +\frac{1}{r}
    \begin{pmatrix}
                  {0} & {-i\partial _{\theta }+eW} \\
                  {-i\partial _{\theta }+eW} & {0} \\
              \end{pmatrix}
\right]
 \begin{pmatrix}
    {\Phi_{+}F_{+}} \\
    {\Phi_{-}F_{-}} \\
  %\end{array}\right)
  \end{pmatrix}
=0
\eeq
where the structure of the $\theta$-dependent spinor component
is dictated by the
angular operator $-\frac{i}{r}\partial_{\theta}+ eW(\theta)$. Thus,  we can factorize the angular part by
requiring
\begin{equation}\label{eq20}
  F_{+}(\theta)= F_{-}(\theta)= F(\theta),  \qquad \left[-i\partial _{\theta }+eW(\theta) \right]F=\varepsilon_{\theta}F
\end{equation}
whose  solution is \beq\label{eq20}
  F(\theta)= e^{ i\left[\varepsilon_{\theta } \theta - e \int W(\theta) d\theta\right]}.
\eeq
The periodicity of the total wave function requires that $\Psi
(r,\theta )=\Psi (r,\theta + 2 \pi )$ and gives the quantized
quantities  $\varepsilon_{\theta}$ which will dependent on the
shape of the non-central part of the potential function $
W(\theta)$. The radial part of the wave function can be simplified
by gauging away the spacial part of the vector potential, i.e
$eR(r)$ term, and reduces to
\beq
    \left(
      \begin{array}{cc}
        {m+S+eV -\varepsilon} & {\partial _{r}+\frac{\varepsilon_{\theta}}{r}} \\
        {-\partial _{r}+\frac{\varepsilon_{\theta}}{r}} & {-m-S+eV-\varepsilon} \\
      \end{array}
    \right)
   \left(
     \begin{array}{c}
       {\Phi_{+}} \\
       {\Phi_{-}} \\
     \end{array}
   \right)=0.
\eeq
Making the change of variable $X=\alpha r$ with
$ \alpha^{2}=\left(\varepsilon-eV\right)^{2}-\left(m+S\right)^{2}$,
$\mu_{\pm}^2=(\varepsilon_{\theta}\mp \frac{1}{2})^2$.
After some
algebra we obtain the second order differential equation
\beq
\label{eq100}
 \left[\frac{d^{2}}{dX^{2}}-\frac{\varepsilon_{\theta}(\varepsilon_{\theta}\mp1)}{X^{2}}+1\right]\Phi _{\pm}=0
\eeq
where we assumed
constant potentials $V$ and $S$ in each region of space.
This equation has some common features with the one
associated with Bessel functions. To clarify this statement, let
us write the solution of \eqref{eq100} as
$\Phi_{\pm}(X)=\sqrt{X}F_{\mu_\pm}(X)$ to obtain
\beq \label{eqF}
  \left[\frac{d^{2}}{dX^{2}}+\frac{1}{X}\frac{d}{dX}-\frac{\mu_{\pm}^2}{X^{2}}+1\right]F_{\mu_\pm}=0.
\eeq
The remaining radial equation \eqref{eqF} is solved in terms of
the Bessel functions. Hence, the general solution is a linear
combination of the two independent Bessel functions. To simplify
our task, we choose $W(\theta)$ in such way that $\int
W(\theta) d\theta = g(\theta)$, with $g(\theta)$ being a periodic function of $\theta$ that implies
$ F(\theta)= e^{ i(\varepsilon_{\theta } \theta-e g(\theta))}$. The boundary condition on the total wave function
$\psi(r,\theta)=\psi(r,\theta+2\pi)$ using the fact that $g(\theta)=g(\theta + 2\pi)$ requires that $e^{i(2\pi
\varepsilon_{\theta}-\sigma_3 \pi)}=1$ giving rise to the quantum number
\beq \varepsilon_{\theta}=\frac{k}{2}, \qquad
k=\pm1, \pm3, \pm5,\cdots
\eeq
which will be denoted
%in the following we denote
$\varepsilon_{\theta}=j$, with $j$ being a half integer. Finally, the most general solution is given by %of our problem reads
\beq \Psi_{\pm} (r, \theta)=\sum_{k,\pm}
\sqrt{\alpha}e^{\frac{i}{2}(k - \sigma _{3}) \theta } \left[A^{'}_{\pm}
J_{\mu_\pm}(\alpha r) +B^{'}_{\pm} Y_{\mu_\pm}(\alpha r) \right].
\eeq
In the following sections we will apply these results to graphene where we set our units such that
$\hbar v_F = 1$ and consider the presence of an induced mass term $m$.

%%%%%%%%%%%%%%%%%%%%%%%%%%%%%%%%%%%%%%%%%%%%%%%%%%%
\section{Graphene quantum dot}
%%%%%%%%%%%%%%%%%%%%%%%%%%%%%%%%%%%%%%%%%%%%%%%

We use our previous work \cite{paper1} and the above formalism to investigate an
interesting case study in graphene. Indeed, we first show how can our findings model
an isolated quantum dot and then study the physical properties of this quantum dot
which might lead to interesting applications.

%%%%%%%%%%%%%%%%%%%%%%%%%%%%%%%%%%%%%%%%%%%%%%%%%%%
\subsection{Isolated quantum dot}
%%%%%%%%%%%%%%%%%%%%%%%%%%%%%%%%%%%%%%%%%%%%%%%

%In this section
Here we consider graphene in the presence of a constant
mass term $m$, that induces a gap of $2m$, which can be realized by
different experimental methods in graphene systems \cite{Zhou,
Giovann}. %In the following we consider the second spin symmetric
%configuration $(V=S)$.
To realize a quantum dot in graphene we consider a cylindrically
symmetric potentials associated with vector potential
$V(r)=V_0\Theta(R-r)$ and pseudo scalar potential
$S(r)=S_0\Theta(R-r)$, where $S_0$ and $V_0$ are  real constants,
$R$ represent the radius of the QD. We would like to study the
potential existence of bound states within the quantum dot. The
solutions of the Dirac equation, that describe the electronic
states inside and outside the quantum dot, are given in terms of
the Bessel function of first and second kind $J_{\mu_{\pm}}(x)$,
$Y_{\mu_{\pm}}(x)$, the modified Bessel function
$I_{\mu_{\pm}}(x)$, $K_{\mu_{\pm}}(x)$ and the Hankel function of
first and second kind $H^{(1,2)}_{\mu_{\pm}}(x)$. These solutions
will be propagating if the wave vectors are real and a decaying
(exponentially decaying) solutions if the wave vectors are
imaginary. We summarize our findings in Table \ref{tab.11} where
we show different domains, which are chosen according to whether
$\alpha$ (for $r < R$) and $\alpha'$ (for $r>R$) are purely
imaginary or real and give, for each case, the suitable Bessel
functions that describe the radial part of the electronic wave
function inside and outside the QD.

\begin{table}[h]
\begin{center}
\begin{tabular}{|c|c|c|}
  \hline
 Domain  &Inside the QD $r<R$   &  Outside the QD $r>R$\\
  \hline
  \hline
  I & $\alpha=\sqrt{(\varepsilon-V_0)^2-(m+S_0)^2}$ & $\alpha'=\sqrt{\varepsilon^2-m^2}$  \\
  & $\Phi(r)\propto J_{|j\mp \frac{1}{2}|}(\alpha r)$ & $\Phi(r)\propto H^{(1,2)}_{|j\mp \frac{1}{2}|}(\alpha' r)$ \\
  %& $\alpha=\sqrt{(\varepsilon-(m+2S_0))(m+\varepsilon)}$ & $\alpha'=\sqrt{\varepsilon^2-m^2}$  \\
   \hline
   \hline
  II & $\alpha=\sqrt{(\varepsilon-V_0)^2 -(m+S_0)^2}$ &  $\eta=\sqrt{m^2-\varepsilon^2}$, \quad  $\alpha'=i\eta$  \\
    & $\Phi(r)\propto J_{|j\mp \frac{1}{2}|}(\alpha r)$ & $\Phi(r)\propto K_{|j\mp \frac{1}{2}|}(\eta r)$ \\
   \hline
   \hline
    III  &$\zeta=\sqrt{(m+S_0)^2 -(\varepsilon-V_0)^2}$, \quad $\alpha=i\zeta$  & $\alpha'=\sqrt{\varepsilon^2-m^2}$ \\
     & $\Phi(r)\propto I_{|j\mp \frac{1}{2}|}(\zeta r)$ & $\Phi(r)\propto H^{(1,2)}_{|j\mp \frac{1}{2}|}(\alpha' r)$ \\
   \hline
   \hline
   IV   & $\zeta=\sqrt{(m+S_0)^2 -(\varepsilon-V_0)^2}$, \quad$\alpha=i\zeta$  &  $\eta=\sqrt{m^2-\varepsilon^2}$, \quad $\alpha'=i\eta$  \\
     & $\Phi(r)\propto I_{|j\mp \frac{1}{2}|}(\zeta r)$ & $\Phi(r)\propto K_{|j\mp \frac{1}{2}|}(\eta r)$ \\
  \hline
\end{tabular}
\end{center}
\caption{\sf{Summarizes  different domains according to
the choice of the wave vectors and their corresponding wave functions.
}}\label{tab.11}
\end{table}
\noindent We note that inside and outside the dot, the wave vectors are
$\alpha^2=(\varepsilon-V)^2-(m+S)^2$ and
$\alpha'^2=\varepsilon^2-m^2$, respectively.
In domain I, one has
%$m+2S_0<\varepsilon$
$m+S_0<|\varepsilon -V_0|$ and $m<\varepsilon$ so that both wave
vectors $\alpha$ and $\alpha'$ are real, the wave function
oscillates inside and outside the QD.
 In domain II,
%$m+2S_0<\varepsilon$
$m+S_0<|\varepsilon -V_0|$ and $m>\varepsilon$, the wave
vector $\alpha$ is real and $\alpha'$ is purely imaginary. In this
region, we have true bound states that oscillate inside the QD and
decay outside. In domain III, we have
 %$m+2S_0>\varepsilon$
$m+S_0>|\varepsilon -V_0|$ and $m<\varepsilon$ that imply $\alpha$
is purely imaginary and $\alpha'$ is real, this gives rise to the tunneling
regime,  that is the wave function decays inside and oscillates
outside the QD. Domain IV is characterized by %$m+2S_0>\varepsilon$
$m+S_0>|\varepsilon -V_0|$ and $m>\varepsilon$ that give both
$\alpha$ and $\alpha'$ purely imaginary and hence the wave
function decays inside and outside the QD. In domain II, the
generale solution of the radial Dirac equation that are regular at
the origin and which decay exponentially as $r\rightarrow \infty$,
are given in term of bessel function $A_{\pm}J_{|j \mp
\frac{1}{2}|}(\alpha r)$ inside the QD and $B_{\pm}K_{|j \mp
\frac{1}{2}|}( \eta r)$ outside the QD. We note that the two other
functions diverges ($Y_\mu(x)$ for $r\rightarrow 0$ and $I_\mu(x)$
for $r\rightarrow \infty$).

The general solutions of Dirac equation, taking a positive value of $j$,  are given by
\bqr\label{eq30}
   \Psi(r,\theta)= \left\{\begin{array}{lll} {\sum_{j} A_{\pm} J_{j\mp \frac{1}{2}}(\alpha r)e^{i(j \mp \frac{1}{2} )\theta},} && {r<R} \\ {\sum_{j} B_{\pm} K_{j\mp
\frac{1}{2}}(\eta r)e^{i(j \mp \frac{1}{2} )\theta},} && {r>R}.
\end{array}\right.
\eqr
The ratios $\frac{A_+}{A_-}$ and $\frac{B_+}{B_-}$ are fixed
by the Dirac equation and the matching conditions of the spinors
at the boundary $r=R$. After a lengthy but straightforward algebra %using the matching condition at $r=R$
we find
\beq\label{ratio}
 \frac{A_+}{A_-}=\frac{\sqrt{\varepsilon-V_0 +(m+S_0)}}{\sqrt{\varepsilon-V_0 -(m+S_0)}}, \qquad  \frac{B_+}{B_-}=\frac{\sqrt{m+\varepsilon}}{\sqrt{m-\varepsilon}}.
\eeq
%\beq\label{ratio}
 %\frac{A_+}{A_-}=\frac{\sqrt{m+\varepsilon}}{\sqrt{\varepsilon-(m+2S_0)}}  \qquad ; \qquad  \frac{B_+}{B_-}=\frac{\sqrt{m+\varepsilon}}{\sqrt{m-\varepsilon}}
%\eeq
Then we obtain the following characteristic equation of the QD
\beq \label{bs} \xi_{+} J_{j- \frac{1}{2}}(\alpha R) K_{j+
\frac{1}{2}}(\eta R)- \xi_{-} J_{j+ \frac{1}{2}}(\alpha R) K_{j-
\frac{1}{2}}(\eta R)=0
\eeq
where $\xi_{\pm}=\sqrt{\left[\varepsilon-V_0
\pm (m+S_0)\right] \left(m\mp S_0\right)}$.
% \beq \label{bs}
%\sqrt{m-\varepsilon} J_{j- \frac{1}{2}}(\alpha R) K_{j+
%\frac{1}{2}}(\eta R)- \sqrt{\varepsilon-(m+2S_0)} J_{j+
%\frac{1}{2}}(\alpha R) K_{j- \frac{1}{2}}(\eta Rj)=0 \eeq
We can see that  \eqref{bs} is symmetric with respect to
the change of the sign of $j$. Thus, we note that for the other
valley $K'$ the corresponding equation is obtained by replacing $j$
by $-j$ in  \eqref{bs}. Therefore below we consider only
positive values of $j$, i.e. $j=\frac{1}{2}, \frac{3}{2}, \frac{5}{2},
\cdots$.\\

\begin{figure}[h]
\centering
  \includegraphics[width=8cm, height=8.5cm]{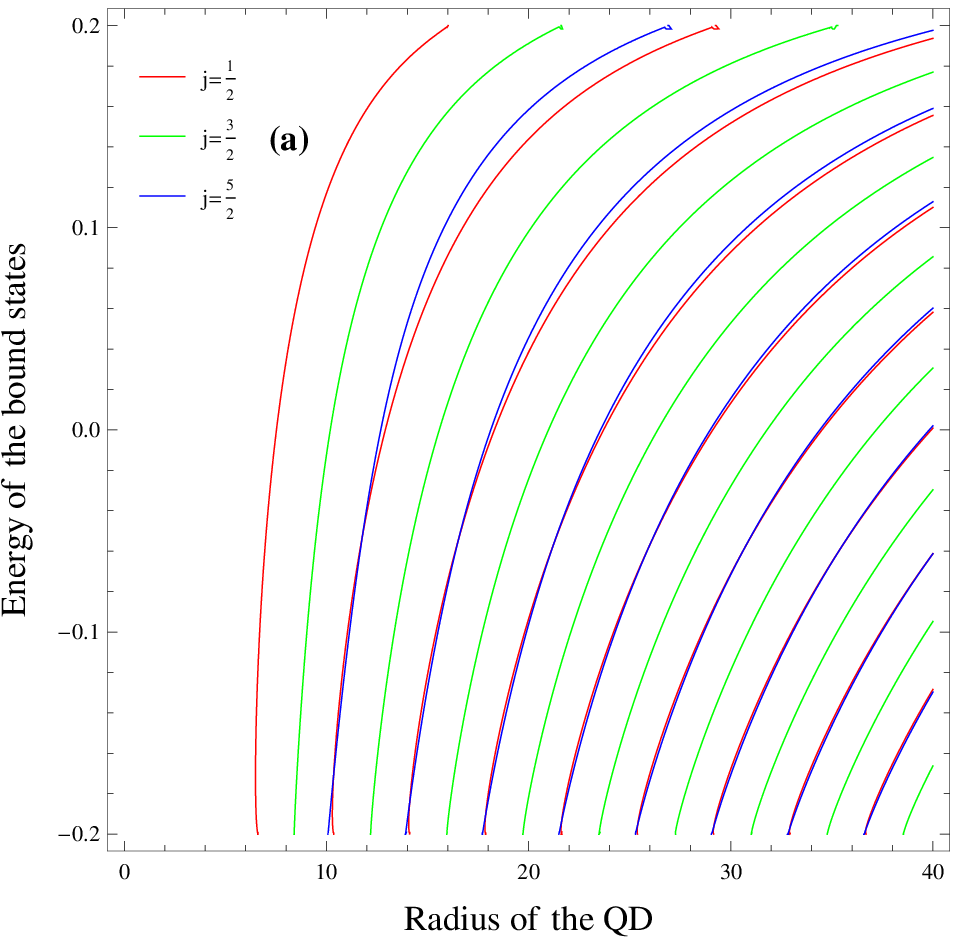}
  \ \ \ \  \includegraphics[width=8cm, height=8.5cm]{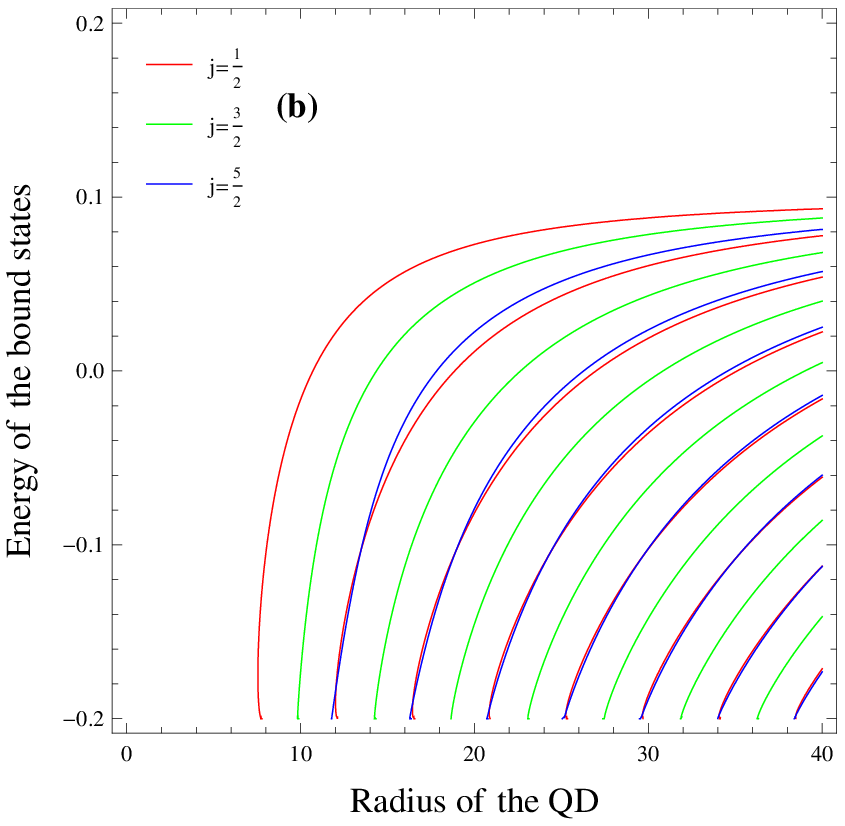}
 \caption{\sf {The contour plot of the energy of the bound states in a graphene quantum dot as a function of the radius $R$ of
 the quantum dot,
  with $j=\frac{1}{2}$, $j=\frac{3}{2}$, $j=\frac{5}{2}$, $m=0.2$, $V_0=0.8$ and $S_0=0.35$ for  (a), $S_0=0.5$ for  (b).}}\label{fig.ste}
\end{figure}
 In Figure \ref{fig.ste} we show the contour plot of the energy of
the bound state in graphene QD as function of the radius of the
QD, for three different values of the half integer $j$ ($\frac{1}{2}$, $\frac{3}{2}$,
$\frac{5}{2}$) and for the strength of the mass term $m=0.2$ and vector potential
$V_0=0.8$. We evaluate the characteristic
equation \eqref{bs} for two different values of the pseudo scalar
potential $S_0$, ($S_0=0.35$ and $S_0=0.5$). We note that if we
consider both valleys $K$ and $K'$ we find that each bound state
is doubly degenerate. From Figure \ref{fig.ste}b,
we can clearly see the effect of the pseudo scalar potential, in fact when we
increase the value of the pseudo scalar potential the number of
the bounds states decreases. We also show that when we increase the
radius of the dot the number of the bounds states increases mainly
for large $R$. Finally, we observe that %From Figure \ref{fig.ste} we can clearly see
%that
more bound states occur in the QD when its radius increases.

 \begin{figure}[h!]
\centering
  \includegraphics[width=8cm, height=8.5cm]{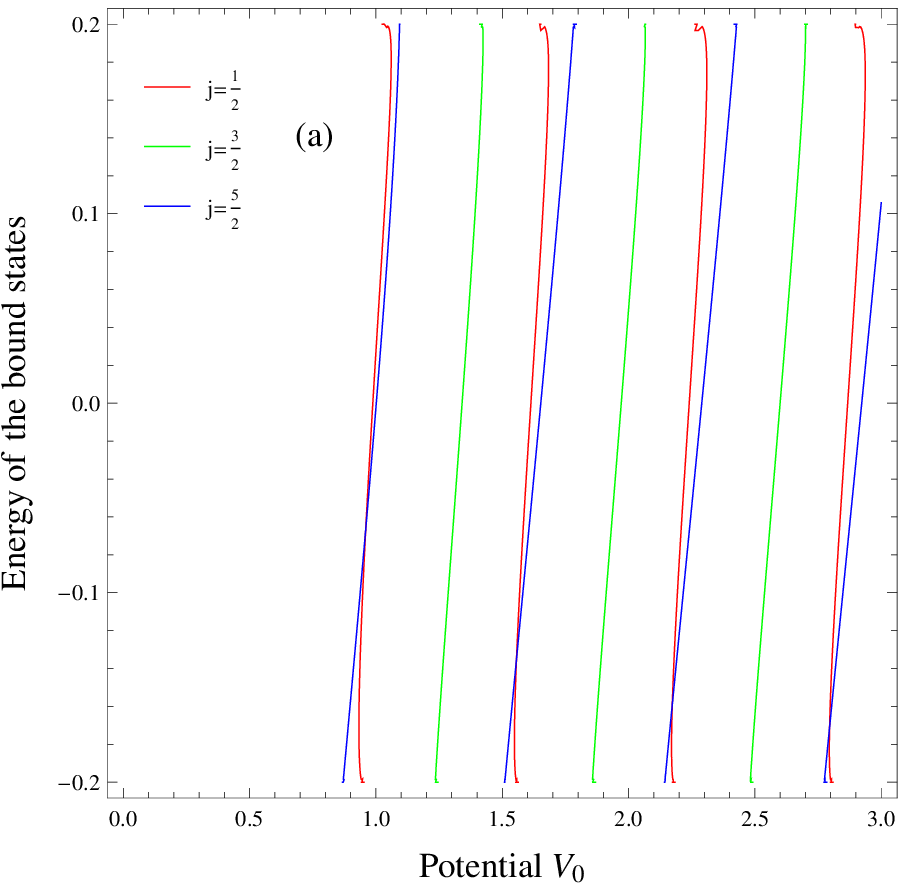}
  \ \ \includegraphics[width=8cm, height=8.5cm]{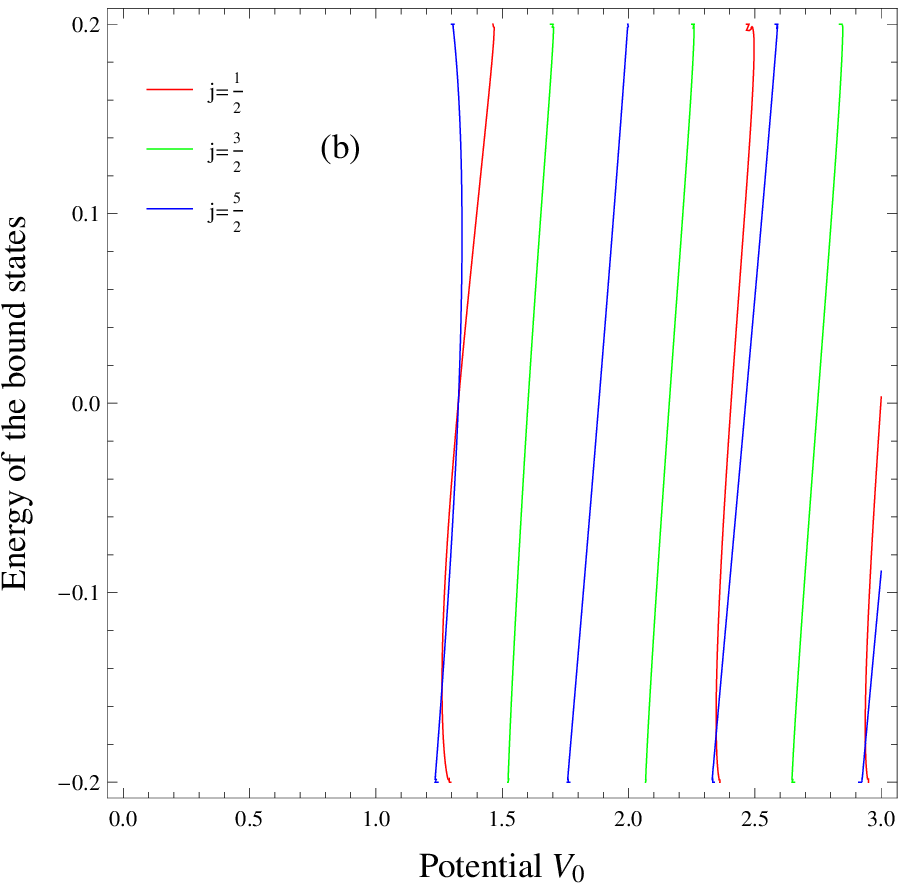}
 \caption{\sf {The contour plot of the energy of the bound states in a graphene quantum dot as a function of the potential
 $V_0$,
  with $j=\frac{1}{2}$, $j=\frac{3}{2}$, $j=\frac{5}{2}$, $m=0.2$, $R=5$ and $S_0=0.1$ for  (a), $S_0=0.8$ for  (b).}}\label{fig.ste_2}
\end{figure}

Figure \ref{fig.ste_2} shows the contour plot of the energy of
 the bound state in graphene QD as a function of the potential strength $V_0$
 for three value of $j$ $\left(\frac{1}{2}, \frac{3}{2},
 \frac{5}{2}\right)$, the mass term used is $m=0.2$ and the radius of the
 QD is taken to be $R=5$. To see the effect of the pseudo scalar potential
 we used two different values $S_0=0.1$ in Figure \ref{fig.ste_2}a
  and $S_0=0.8$ in Figure \ref{fig.ste_2}b.

From the above computations we note that more bound states
can be accommodated in the QD when the strength of
the confining potential or the radius of the QD increase. These results are in
agreement with previous work on bound state energies in
radially symmetric graphene QD \cite{Pal, Nilsson}. In Figure \ref{fig.ste-RV}
we show six bound states for $V_0=0.6$ and $R=30$: two for $j=\frac{1}{2}$, two for
$j=\frac{3}{2}$ and two for $j=\frac{5}{2}$.\\

\begin{figure}[h!]
\centering
  \includegraphics[width=8cm, height=8.5cm]{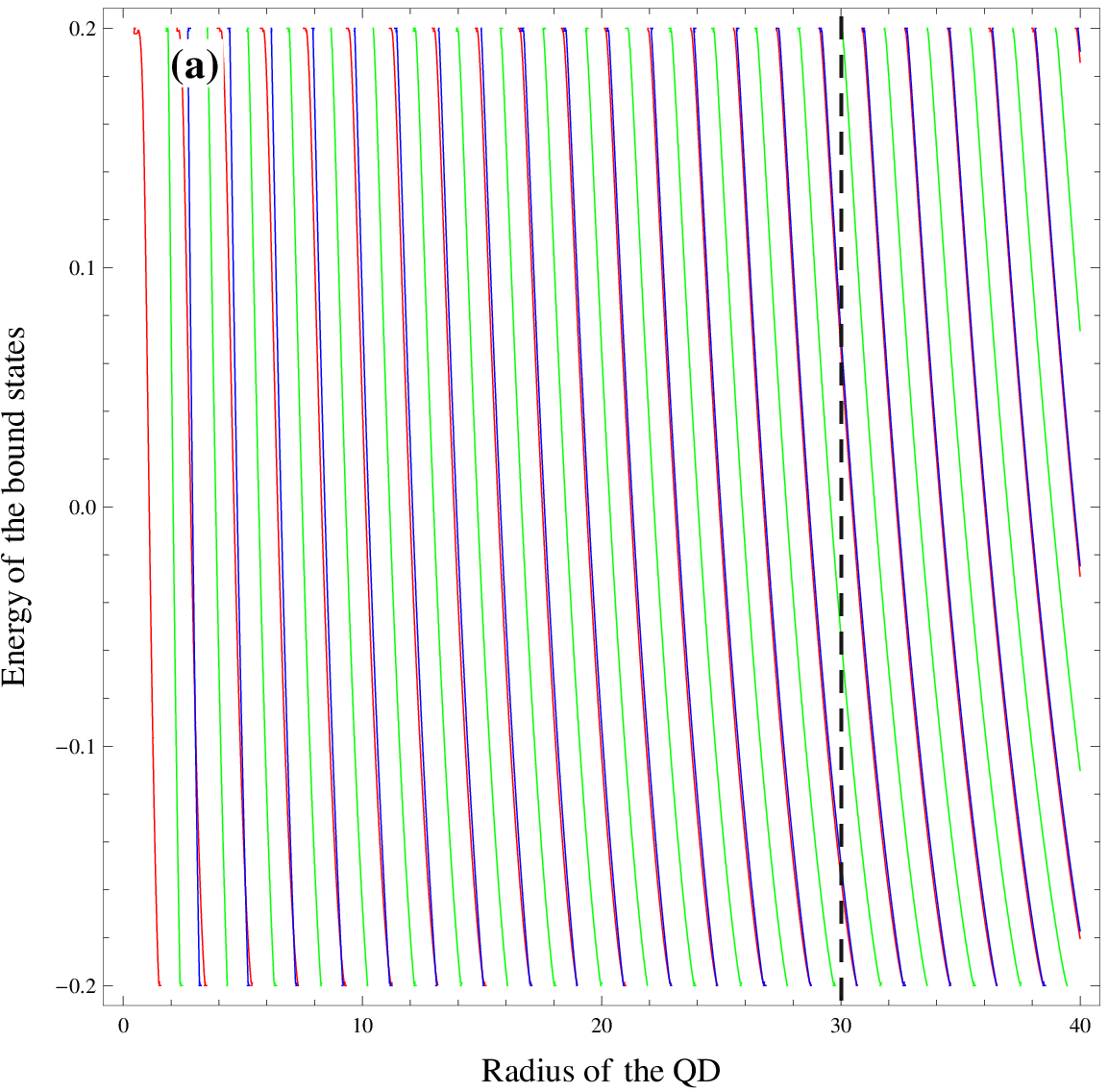}
  \ \ \includegraphics[width=8cm, height=8.5cm]{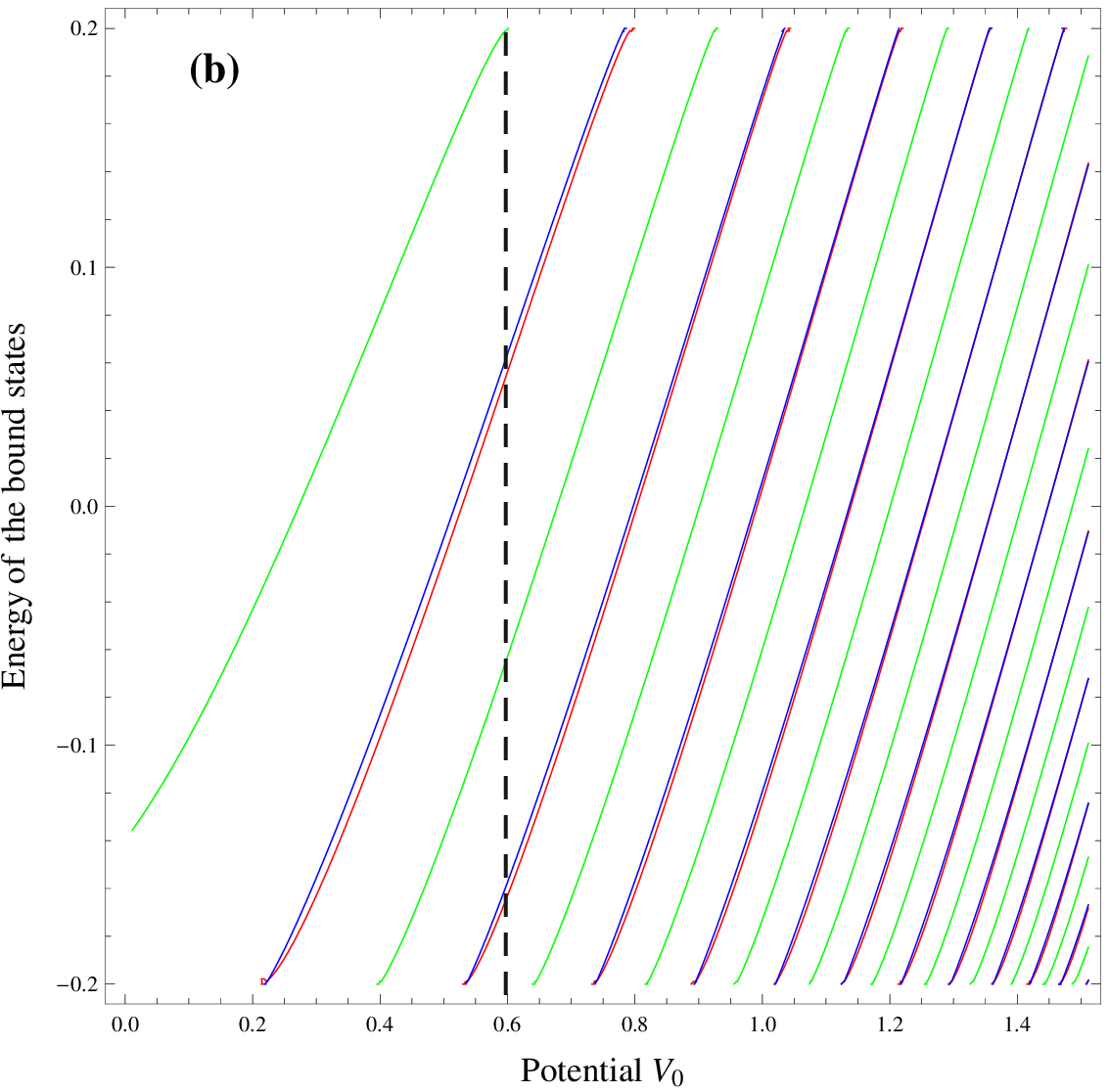}
 \caption{\sf {(a): The contour plot of the energy of the bound states in a graphene quantum dot as a function of the
 radius $R$. (b): The contour plot of the energy of the bound states in a graphene quantum dot as a function of the potential
 $V_0$.
 With $j=\frac{1}{2}$ (red line), $j=\frac{3}{2}$ (green line),
   $j=\frac{5}{2}$ (blue line), $m=0.2$, and $S_0=1.6$. $V_0=0.6$ for (a), $R=30$ for  (b).}}\label{fig.ste-RV}
\end{figure}

\begin{figure}[H]
\centering
  \includegraphics[width=8cm, height=7.5cm]{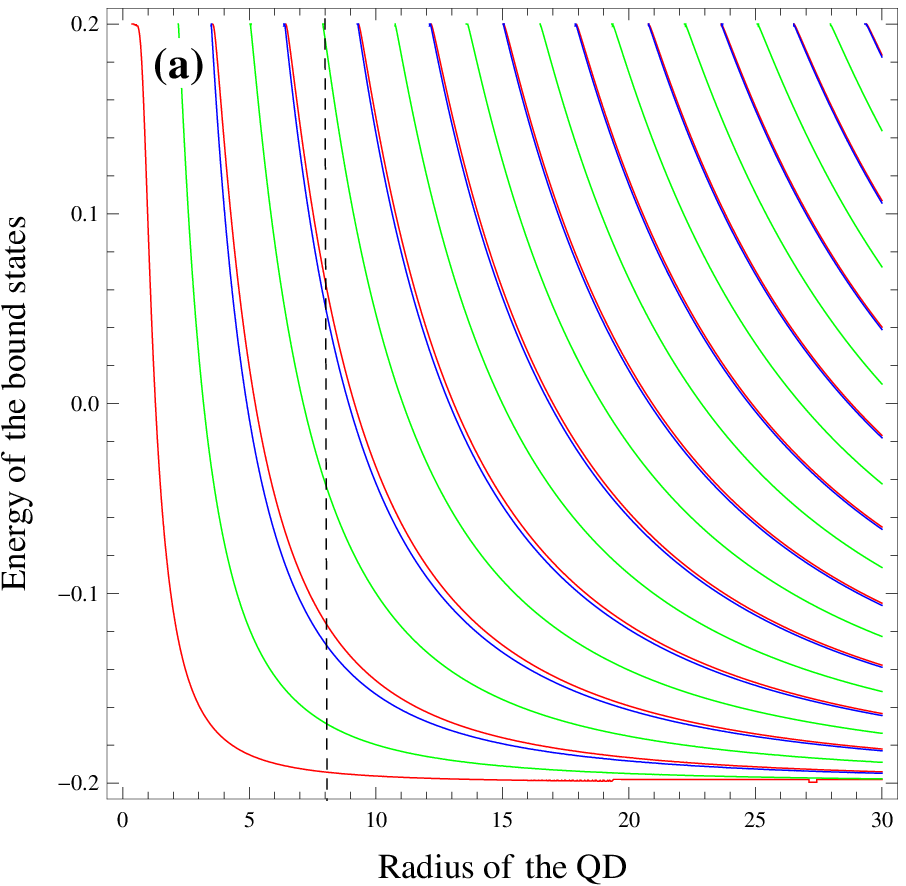}
  \ \ \includegraphics[width=8cm, height=7.5cm]{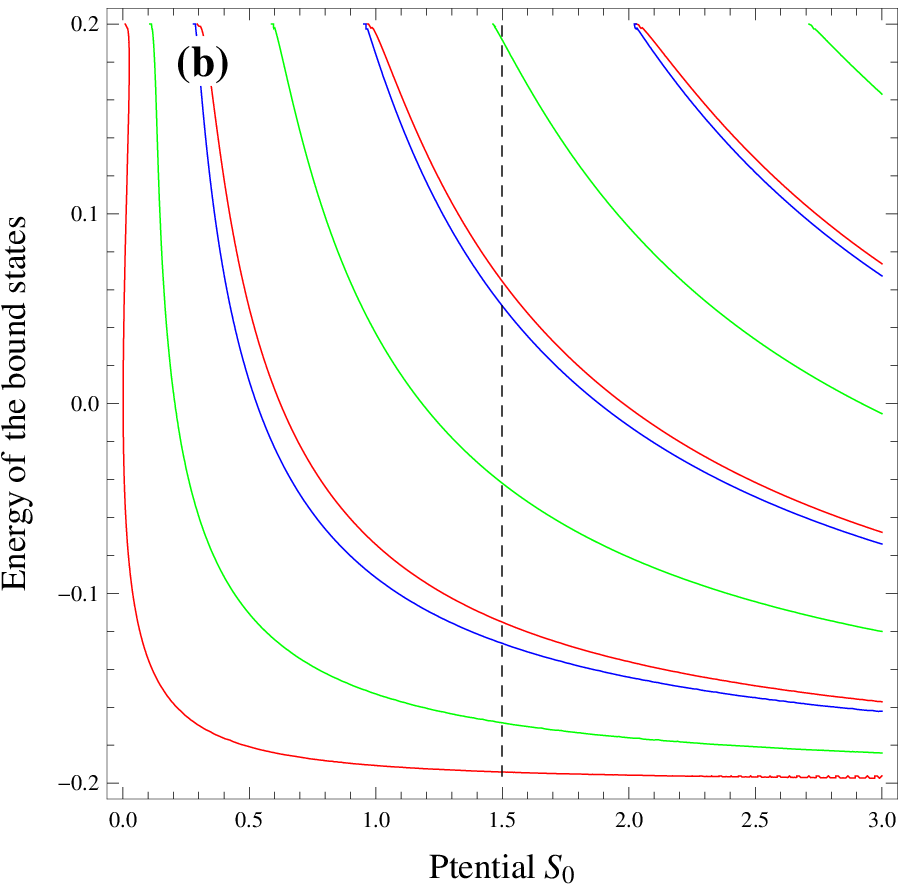}
 \caption{\sf {(a): The contour plot of the energy of the bound states in a graphene quantum dot as a function of the
 radius $R$. (b): The contour plot of the energy of the bound states in a graphene quantum dot as a function of the potential
 $S_0$.
  With $j=\frac{1}{2}$ (red line), $j=\frac{3}{2}$ (green line),
   $j=\frac{5}{2}$ (blue line), $m=0.2$, and $S_0=1.5$ for
   (a), $R=8$ for
    (b).}}\label{fig.ste-RS2}
\end{figure}

At this stage, let us  consider the case of spin symmetric configuration with
$V = S$. Within the framework of the Dirac equation, this configuration occurs when
the difference between the Lorentz vector potential $V(r)$ and the Lorentz scalar potential $S(r)$ is constant, that is,
$\Delta(r) = V(r) - S(r)= \mbox{constant}$, which in our case this constant is taken to be zero. The near experimental realization
of the spin symmetric potential configuration may explain the degeneracy in some heavy meson spectra \cite{Meson}.
In this case  \eqref{bs} becomes
  \beq \label{bs2}
\sqrt{m-\varepsilon} J_{j- \frac{1}{2}}(\alpha R) K_{j+
\frac{1}{2}}(\eta R)- \sqrt{\varepsilon-(m+2S_0)} J_{j+
\frac{1}{2}}(\alpha R) K_{j- \frac{1}{2}}(\eta Rj)=0. \eeq
Similarly to what we have seen in \eqref{bs}, \eqref{bs2} is also symmetric with
respect to the transformation $j\rightarrow -j$.

In Figure \ref{fig.ste-RS2} we show the solution of the equation
\eqref{bs2}. We note that when $R=8$ and $S_0=1.6$ we have eight
bound states: three for $j=\frac{1}{2}$, three for $j=\frac{3}{2}$
and two for $j=\frac{5}{2}$. We can also show that when we
increase the radius of the dot or the strength of the pseudo
scalar potential we accommodate more bound states.

%%%%%%%%%%%%%%%%%%%%%%%%%%%%%%%%%%%%%%%%%%%%%%%%%%%%%%%%%%%
\subsection{Specific potential configuration}
%%%%%%%%%%%%%%%%%%%%%%%%%%%%%%%%%%%%%%%%%%%%%%%%%%%%%%%%%%%

%In this subsection
We consider the first configuration discussed in
section 3 and study graphene in the presence of a mass term
$m$ that induces a gap of value $2m$. % which can be realized by different
%experimental methods \cite{zhou, Giovann}.
The solution of the Dirac equation in this case is given in terms of the Whittaker hypergeometric
functions $M_{\nu,\mu_{\pm}}(2\gamma r)$ and $W_{\nu,\mu_{\pm}}(2\gamma r)$, such as
%which can be expressed in terms of the confluent hypergeometric function
%$_{1}F_{1}$ and $U$ \cite{Abramowitz}, such as
%. Thus,
%the general solution reads as %of the Dirac equation in this case reads
 \beq \Psi (r, \theta)=\sum_{k,\pm}
\frac{1}{\sqrt{r}}e^{\frac{i}{2}(k - \sigma _{3}) \theta }[{\rm
{\mathcal A_{\pm}}} M_{\nu,\mu_{\pm}}(2\gamma r) +{\rm {\mathcal
B_{\pm}}} W_{\nu,\mu_{\pm}}(2\gamma r)]
\eeq
where
$M_{\nu,\mu_{\pm}}(2\gamma r)$ and $W_{\nu,\mu_{\pm}}(2\gamma r)$ are given in
terms of confluent hypergeometric functions \cite{Abramowitz}
\bqr
&& M_{\nu,\mu_{\pm}}(2\gamma r)= e^{-\gamma r} (2\gamma r)^{{\mu_{\pm}} +1/2} {}_1F_1\left(\frac{1}{2} +\mu_{\pm} - \nu,1+2\mu_{\pm},2\gamma r\right)\\
&& W_{\nu,\mu_{\pm}}(2\gamma r)= e^{-\gamma r} (2\gamma r)^{\mu_{\pm} +1/2}
U\left(\frac{1}{2} +\mu_{\pm} - \nu,1+2\mu_{\pm},2\gamma r\right). \eqr
Then, the general solution, that is regular at the
origin and decays exponentially at large $r$, can be written in terms of the
confluent hypergeometric function as
\beq\label{eq30}
    (2\gamma)^{\mu_{\pm}+\frac{1}{2}} e^{-\gamma r} r^{\mu_{\pm}} \left\{\begin{array}{lll} {{\rm {\mathcal A_{\pm}}} \  {}_{1}F_{1}\left(\frac{1}{2}+\mu_{\pm}-\nu,1+2\mu_{\pm},2 \gamma r\right),} && {r<R} \\
   {{\rm {\mathcal B_{\pm}}}\
   U\left(\frac{1}{2}+\mu_{\pm}-\nu,1+2\mu_{\pm},2 \gamma r\right),} && {r>R}
\end{array}\right.
\eeq
where $\mu_{\pm}=j\mp \frac{1}{2}$, $\nu=-\frac{W}{\gamma}j$ and
$\gamma^{2}=\left(m+S\right)^{2}+W^2-\left(\varepsilon-eV\right)^{2}$.

At this stage let us investigate the bound states. For this purpose, we adopt
%In order to find the bound states, we consider the following
the explicit potential configuration
\beq\label{eq30}
    V(r)=\left\{\begin{array}{lll} {V_{0}} && {r< R} \\ {0} && {r>R} \end{array}\right., \qquad
    W(r)=\left\{\begin{array}{lll} {W_{0}} && {r<R} \\ {0} && {r> R} \end{array}\right., \qquad
    S(r)=\left\{\begin{array}{lll} {S_{0}} && {r< R} \\ {0} && {r> R} \end{array}\right.
\eeq
which will help us simplify the above established formalism. The two matching conditions  of the spinors at $r=R$ give
 \bqr \label{b.coud}
&& \gamma^{j} e^{-\gamma R} {\rm {\mathcal
A_+}}\ {_{1}F_{1}}(j-\nu,2j,2 \gamma R) =  {\gamma'}^{j}
e^{-\gamma^{'} R} {\rm {\mathcal B_+}}\ U(j,2j,2 \gamma^{'}R) \\
&& \gamma^{j+1} e^{-\gamma R} {\rm {\mathcal
A_-}}\ {_{1}F_{1}}(j+1-\nu,2(j+1),2 \gamma R) = {\gamma'}^{j+1}
e^{-\gamma^{'} R} {\rm {\mathcal B_-}}\ U(j+1,2(j+1),2 \gamma^{'}R).
\eqr
Now inside and outside the QD we have, respectively, the quantities
%we have inside the QD
 $\gamma^{2}=\left(m+S_0 \right)^{2}+W_{0}^2-\left(\varepsilon-V_0\right)^{2}$
%and outside the QD
and $\gamma^{'2}=m^{2}-\varepsilon^{2}$. The condition for the existence of the
bound states can be written as follows %is obtained
\beq \label{eq.boun.s2}\gamma^{'} \frac{{\rm {\mathcal
A_+}}}{{\rm {\mathcal A_-}}} \frac{{_{1}F_{1}}(j-\nu,2j,2 \gamma
R)}{{_{1}F_{1}}(j+1-\nu,2(j+1),2 \gamma R)}=\gamma \frac{{\rm
{\mathcal B_+}}}{{\rm {\mathcal B_-}}} \frac{U(j,2j,2 \gamma^{'}
R)}{U(j+1,2(j+1),2 \gamma^{'} R)}
\eeq
where the ratios $\frac{{\rm
{\mathcal A_+}}}{{\rm {\mathcal A_-}}}$ and $\frac{{\rm {\mathcal
B_+}}}{{\rm {\mathcal B_-}}}$ are fixed by the coupled
differential equation \eqref{eq33} using the general solutions for
$r<R$ and $r>R$. After a straightforward but lengthy algebra we
obtain
\beq \frac{{\rm {\mathcal A_+}}}{{\rm {\mathcal
A_-}}}=-2\frac{\sqrt{(m+S_0)^2+W_{0}^2+(\varepsilon-V_0)^2}}{m+S_0-\varepsilon+V_0}(2j+1), \qquad
 \frac{{\rm {\mathcal B_+}}}{{\rm {\mathcal
B_-}}}=\frac{\sqrt{m+\varepsilon}}{\sqrt{m-\varepsilon}}
\eeq
%we take $W=0$ (i.e $\nu=0$)
and  \eqref{eq.boun.s2} becomes
\beq \label{eq.boun} \xi {_{1}F_{1}}(j-\nu,2j,2 \gamma R)
U(j+1,2(j+1),2 \gamma^{'} R) + \zeta {_{1}F_{1}}(j+1-\nu,2(j+1),2
\gamma R) U(j,2j,2 \gamma^{'} R) =0 \eeq where
$\xi=2(2j+1)(m-\varepsilon)$, $\zeta=(m+S_0 -(\varepsilon-V_0))$,
%$\gamma$ become
$\gamma=\sqrt{(m+S_0)^2+W_{0}^2-(\varepsilon-V_0)^2}$,
$\gamma^{'}=\sqrt{m^2-\varepsilon^2}$ and
$\nu=-\frac{W_0}{\gamma}j$.\\

\begin{figure}[h!]
\centering
  \includegraphics[width=8cm, height=8.5cm]{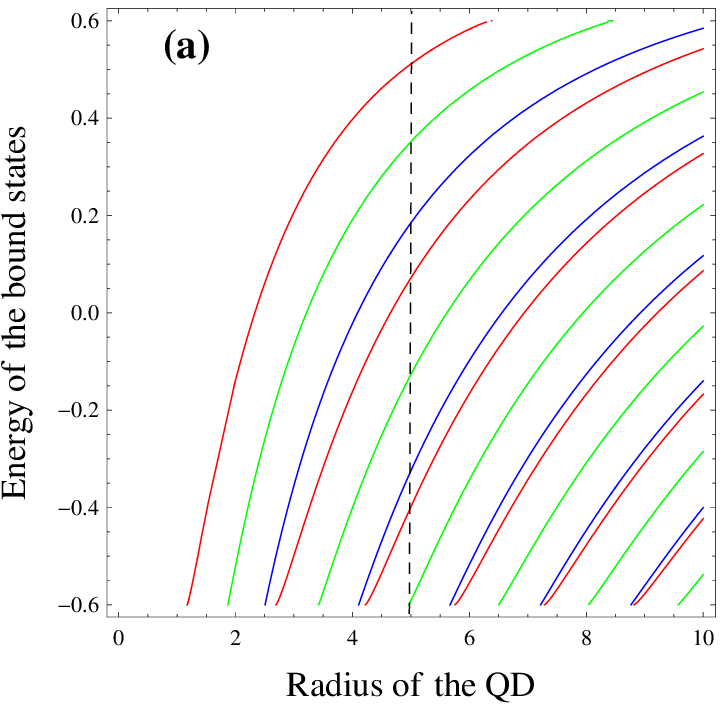}
 \ \ \    \includegraphics[width=8cm, height=8.5cm]{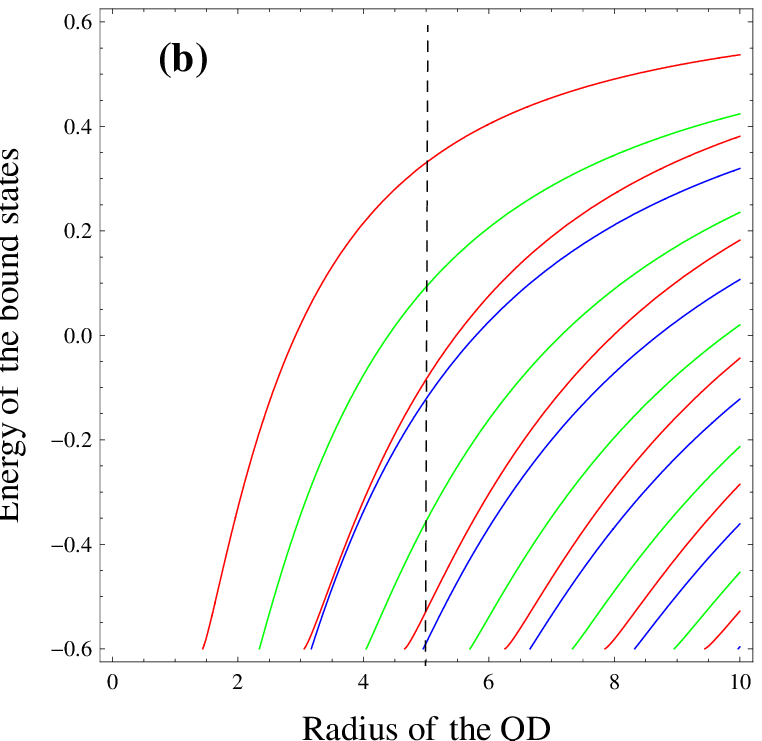}
 \caption{\sf {The contour plot of the energy of the bound states in a graphene quantum dots as a function of the radius $R$ of the quantum dot with $j=\frac{1}{2}$, $j=\frac{3}{2}$, $j=\frac{5}{2}$, $m=0.6$, $S_0=0.2$, $V_0=1.6$ and $W_0=0$ for  (a), $W_0=0.5$ for  (b)}}\label{fig.bd-st2}
\end{figure}
In Figure \ref{fig.bd-st2} we show the energy of the QD as a
function of the dot radius $R$. Evaluating
\ref{eq.boun}, we show the states only for $j=\frac{1}{2},
\frac{3}{2}, \frac{5}{2}$. We use $m=0.6$, $V_0=1.6$, $S_0=0.2$
and $W_0=0$ in Figure \ref{fig.bd-st2}a, $W_0=0.5$ in Figure
\ref{fig.bd-st2}b. %From Figure \ref{fig.bd-st2}
We can clearly see the effect of the pseud-scalar potential, its presence leads
to the reduction of the number of the bound states. In the situation
where $W_0=0$ and $R=5$ we have eight bounds states: three with
$j=\frac{1}{2}$, three with $j=\frac{3}{2}$ and two with
$j=\frac{5}{2}$. On the other hand, for $W_0=0.5$ and for the same
value of $R$ $(R=5)$, we have only seven bounds states: three with
$j=\frac{1}{2}$, two with $j=\frac{3}{2}$ and two with $j=\frac{5}{2}$.
From Figure \ref{fig.bd-st2} it is clearly seen that when we increase the radius of the QD more bound state
can be accommodated in the QD, the same remark was made for the first
potential configuration. The presence of a pseud-scalar potential leads to the
reduction of the number of the bounds states.\\

 \begin{figure}[h!]
\centering
  \includegraphics[width=8cm, height=8.5cm]{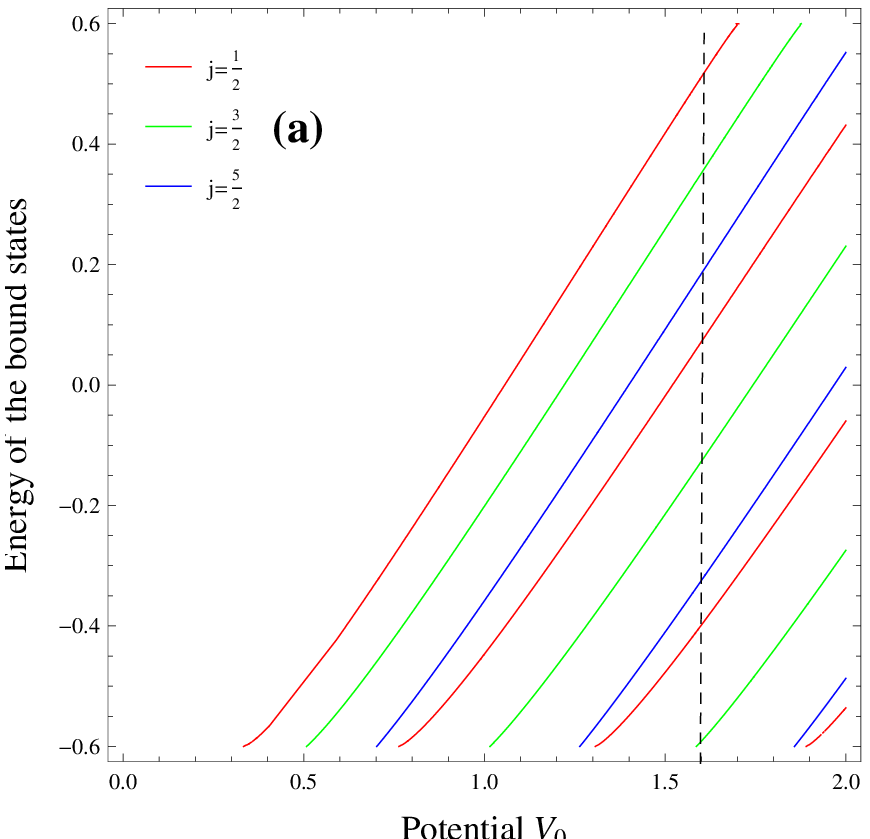}
  \ \ \includegraphics[width=8cm, height=8.5cm]{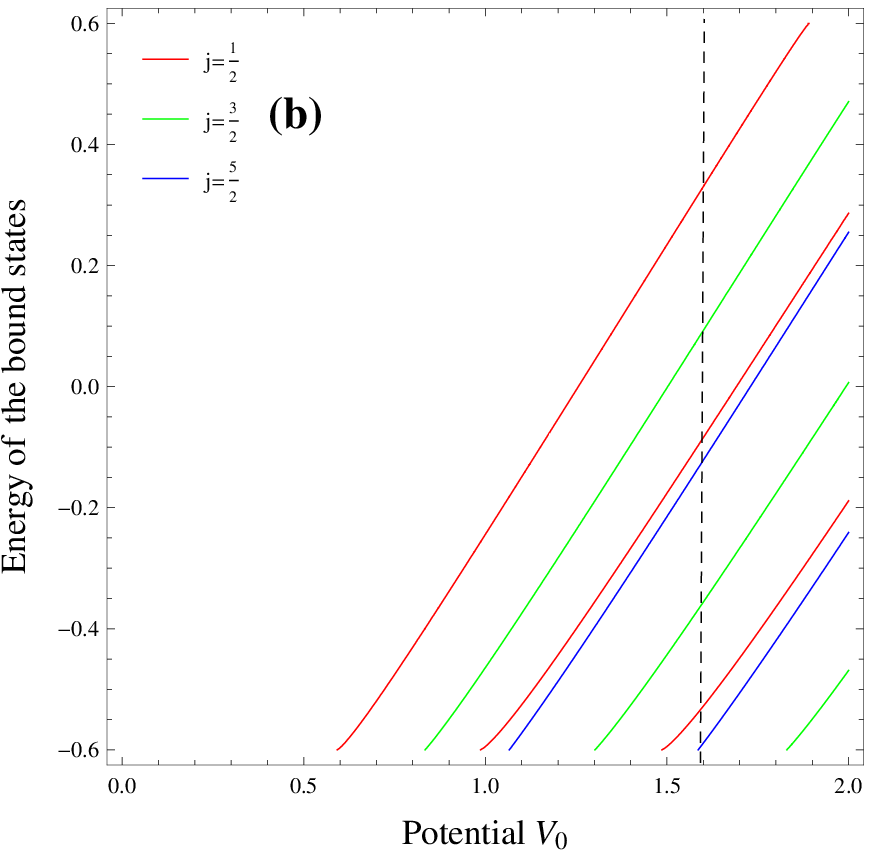}
 \caption{\sf {The contour plot of the energy of the bound states in a graphene quantum dots as a function of the potential $S_0$
  with $j=\frac{1}{2}$, $j=\frac{3}{2}$, $j=\frac{5}{2}$, $m=0.6$, $S_0=0.2$, $R=5$ and $W_0=0$ for  (a), $W_0=0.5$ for  (b).}}\label{fig.ste2.2}
\end{figure}
In Figure \ref{fig.ste2.2} we show the evolution of the energy
of the bound states as function of the strength of the potential $V_0$. Similarly to
Figure \ref{fig.bd-st2}, we present only the states  with
$j=\frac{1}{2}, \frac{3}{2}, \frac{5}{2}$. We use $m=0.6$,
$S_0=0.2$, $R=5$ and $W_0=0$ in Figure \ref{fig.ste2.2}a and
$W_0=0.5$ in Figure \ref{fig.ste2.2}b. We note that when
$V_0=1.6$ and $W_0=0$ we have eight bounds states: three with
$j=\frac{1}{2}$, three with $j=\frac{3}{2}$ and two with
$j=\frac{5}{2}$. On the other hand
when $W_0=0.5$, for the same $V_0=1.6$ we only have seven bounds states: three with
$j=\frac{1}{2}$, two with $j=\frac{3}{2}$ and two with
$j=\frac{5}{2}$. These results are is in good agreement with Figure
\ref{fig.bd-st2}.
\noindent
It is clearly seen from Figure \ref{fig.ste2.2} that when we
increase the radius of the QD we can accommodate more bound states
like in the first case. In conclusion, the presence of a pseud-scalar
potential leads to the reduction of the number of the bound
states.

%%%%%%%%%%%%%%%%%%%%%%%%%%%%%%%%%%%%%%%%%%%%%%%%%%%%%%%%%%%%%%%%%%%%%
\section{Transport properties of graphene quantum dot}
%%%%%%%%%%%%%%%%%%%%%%%%%%%%%%%%%%%%%%%%%%%%%%%%%%%%%%%%%%%%%%%%%%%%%

In order to calculate the transport properties through a
circular graphene quantum dot, we consider the spin symmetric potential
configuration with $V=S$ %. Using the following potential:
and use a radially symmetric potential profile
\beq \label{potential}
S(r)=\left\{\begin{array}{lll}
{U,}&{}&{r<a} \\
{0,} &{}&{a<r<b}\\
{S,}&{}&{b<r.}
 \end{array}\right.
\eeq
This potential represents a quantum dot of radius $a$ connected to an environment
represented by the external region $r > b$. Bound states of the quantum dot are
allowed to leak through the junction region, $a < r < b$, to reach the environment at $r > b$.
The general solution of the Dirac equation is oscillatory for
$r<a$ and decays exponentially for $b<r$, so that we can have
incoming and outgoing waves in the intermediate region ($a<r<b$)
and define a scattering cross section. The radial solutions of the
Dirac equation are given by
\beq \label{sol} \left\{\begin{array}{lll}
{A_{\pm}J_{\nu_{\pm}}(\beta r),}&{}&{r<a} \\
{ A_{1\pm}K_{\nu_{\pm}}(\eta r)+B_{1\pm}I_{\nu_{\pm}}(\eta r),} &{}&{a<r<b}\\
{A_{2\pm}H^{(1)}_{\nu_{\pm}}(\alpha
r)+B_{2\pm}H^{(2)}_{\nu_{\pm}}(\alpha r),}&{}&{b<r}
 \end{array}\right.
 \eeq
where $\alpha=\sqrt{(\varepsilon - (m+2S))(m+\varepsilon)}$,
$\eta=\sqrt{m^2-\varepsilon^2 }$ and $\beta=\sqrt{(\varepsilon -
(m+2U))(m+\varepsilon)}$.

The Dirac equation, $H\Psi=\varepsilon\Psi$,
has a scattering solution and the corresponding %that has the following
asymptotic form can be written as
\cite{Landau}
\beq
\Psi(r,\theta)=\Psi_{\sf in}(r,\theta)+\Psi_{\sf out}(r,\theta)
\eeq
where
$\Psi_{\sf in}(r,\theta)$ is the incoming plane wave in the $x$-direction
and $\Psi_{\sf out}(r,\theta)$ is the scattered outgoing
spherical wave. These solutions have asymptotic behavior as $r\rightarrow\infty$ given by
\bqr
 && \Psi_{\sf in}(r,\theta)
=\frac{1}{\sqrt{2}}e^{i\alpha r \cos\theta}  \begin{pmatrix}
       {\sqrt{m+\varepsilon}} \\
       {\sqrt{\varepsilon - (m+2S)}} \\
     \end{pmatrix}\\
&& \Psi_{\sf out}(r,\theta) =\frac{e^{i\alpha r}}{\sqrt{ \pi \alpha
r}}f_{\pm}(\theta)\begin{pmatrix}
       { \sqrt{m+\varepsilon}} \\
       { \sqrt{\varepsilon - (m+2S)} } \\
     \end{pmatrix}
\eqr
where $f(\theta)$ defines the scattering amplitude and $\theta$ is the scattering angle.
%Using
Next, we use the decomposition of the
plane wave \beq\label{decompo}
e^{i\alpha x}=\sum \limits_{m=-\infty}^\infty i^m J_m(k_2r) e^{i m\theta}
\eeq
and the asymptotic forms of the Bessel functions at large $r$
% and the factor $-i
%=e^{-i \frac{\pi}{4}}$ is introduced for further convenience.
\bqr
\label{j}
&& J_{m}(\alpha r)\approx \frac{1}{\sqrt{2 \pi \alpha
r}}  \left( e^{i(\alpha r
-\frac{m}{2}\pi-\frac{\pi}{4})}+e^{-i(\alpha r
-\frac{m}{2}\pi-\frac{\pi}{4})} \right) \\
 &&
\label{h} H^{(1,2)}_{m}(\alpha r)\approx \sqrt{\frac{2}{\pi \alpha
r}} e^{\pm i \left( \alpha r -\frac{m}{2}\pi-\frac{\pi}{4}
\right)}.
\eqr

In order to extract more information about the present system, let us
study the differential and total cross sections. These  are given by
\cite{Landau}
\beq \frac{d \Lambda(\theta)}{d
\theta}=\frac{|f(\theta)|^2}{2 \pi \alpha}, \qquad
\Lambda=\oint \frac{|f(\theta)|^2}{2 \pi \alpha} d\theta.
\eeq
where the scattering amplitude $f_{\pm}(\theta)$ can be  expanded in Fourier series with coefficient $f_j$,
such as
\beq f_{\pm}(\theta)=\frac{1}{\sqrt{2}} \sum_j
e^{i((j\mp \frac{1}{2})-\frac{\pi}{4})\theta}f_j .\eeq
Combining all these results to write the asymptotic solutions as
\beq \label{phi}
    \Psi_\pm(r,\theta)= \sum \limits_{m=-\infty}^\infty  \frac{i^{m}}{\sqrt{2 \pi \alpha r}}
    \left[\left(1+f_j\right)e^{i(\alpha r-\frac{m \pi}{2}-\frac{\pi}{4})}+e^{-i(\alpha
    r-\frac{m
    \pi}{2}-\frac{\pi}{4})}\right]\begin{pmatrix}
       { \sqrt{m+S}} \\
       { \sqrt{\varepsilon - (m+2S)} } \\
     \end{pmatrix}
      e^{im\theta}.
\eeq
Turning back to the solution in \eqref{sol}, the
asymptotic form for the spinor wave function when $b<r$ can be expressed as
\beq
\label{sol-j} \Phi_{j}(r)\approx \sqrt{\frac{2}{\pi \alpha r}}
\left( A_{2\pm} e^{i(\alpha r -\frac{m}{2}\pi-\frac{\pi}{4})} +
B_{2\pm} e^{-i(\alpha r -\frac{m}{2}\pi-\frac{\pi}{4})}\right)
\eeq
where the parameters are defined by
\bqr
&& A_{2+}=\sqrt{m+\varepsilon}  S_j,  \qquad
A_{2-}=\sqrt{\varepsilon-(m+2S)} S_j  \\
&& B_{2+}=\sqrt{m+\varepsilon}, \qquad
B_{2-}=\sqrt{\varepsilon-(m+2S)}.
\eqr
We have defined the scattering matrix $S_j$ by $S_j= f_j+1$. Applying the
matching conditions at the boundaries $r=a$ and $ r=b$ we obtain
\begin{eqnarray} \label{GEQ129}
\begin{array}{lll}
{A_{+} J_{j-\frac{1}{2}}(\beta a)}&=& {A_{1+} K_{j-\frac{1}{2}}(\eta a)+ B_{1+} I_{j-\frac{1}{2}}(\eta a)} \nonumber\\
{A_{-} J_{j+\frac{1}{2}}(\beta a)}&=&  {A_{1-} K_{j+\frac{1}{2}}(\eta a)+ B_{1-} I_{j+\frac{1}{2}}(\eta a)}\\
{A_{1+} K_{j-\frac{1}{2}}(\eta b)+ B_{1+} I_{j-\frac{1}{2}}(\eta b)}&=&{A_{2+} H^{(1)}_{j-\frac{1}{2}}(\alpha b)+ B_{2+} H^{(2)}_{j-\frac{1}{2}}(\alpha b)} \nonumber \\
{A_{1-} K_{j+\frac{1}{2}}(\eta b)+ B_{1-} I_{j+\frac{1}{2}}(\eta
b)}&=&{A_{2-} H^{(1)}_{j+\frac{1}{2}}(\alpha b)+ B_{2-}
H^{(2)}_{j+\frac{1}{2}}(\alpha b)} \nonumber
\end{array}
\end{eqnarray}
The relationship between the different coefficients reads
\beq
\frac{A_{1+}}{A_{1-}}=\frac{\sqrt{m+\varepsilon}}{\sqrt{m-\varepsilon}}, \qquad
\frac{B_{1+}}{B_{1-}}=-\frac{\sqrt{m+\varepsilon}}{\sqrt{m-\varepsilon}}, \qquad
\frac{A_{+}}{A_{-}}=\frac{\sqrt{m+\varepsilon}}{\sqrt{\varepsilon-(m+2
U)}}. \eeq
Defining the matrix
\bqr
&&M^{(1,2)}=\\
&&\begin{pmatrix}
{- \sqrt{m+\varepsilon} J_{j-\frac{1}{2}}(\beta a) }& {\sqrt{m+\varepsilon} K_{j-\frac{1}{2}}(\eta a)} & {\sqrt{m+\varepsilon} I_{j-\frac{1}{2}}(\eta a)} & {0 }\\
 {- \sqrt{\varepsilon-(m+2U)} J_{j+\frac{1}{2}}(\beta a)} & {\sqrt{m-\varepsilon} K_{j+\frac{1}{2}}(\eta a) }&{-\sqrt{m-\varepsilon} I_{j+\frac{1}{2}}(\eta a) }& {0} \\
 {0 }& \sqrt{m+\varepsilon} K_{j-\frac{1}{2}}(\eta b) &{ \sqrt{m+\varepsilon}I_{j-\frac{1}{2}}(\eta b)}& {\sqrt{m+\varepsilon}H^{(1,2)}_{j-\frac{1}{2}}(\alpha b)} \\
 {0} & \sqrt{m-\varepsilon} K_{j+\frac{1}{2}}(\eta b) & {-\sqrt{m-\varepsilon}I_{j+\frac{1}{2}}(\eta b) }&{ \sqrt{\varepsilon-(m+2S)}H^{(1,2)}_{j+\frac{1}{2}}(\alpha b) }
\end{pmatrix} \nonumber
\eqr
After a straightforward algebra we obtain the scattering matrix as
\beq S_j=-\frac{\det M^2}{\det M^1}.
\eeq

%and the transport cross section is defined as \cite{katnelson}
%\beq \sigma _{tr}=\int_{- \pi}^{\pi} d
%\theta(1-\cos{\theta})\frac{d \Lambda(\theta)}{d\theta} \eeq

\begin{figure}[h]
\centering
  \includegraphics[width=8cm, height=6.5cm]{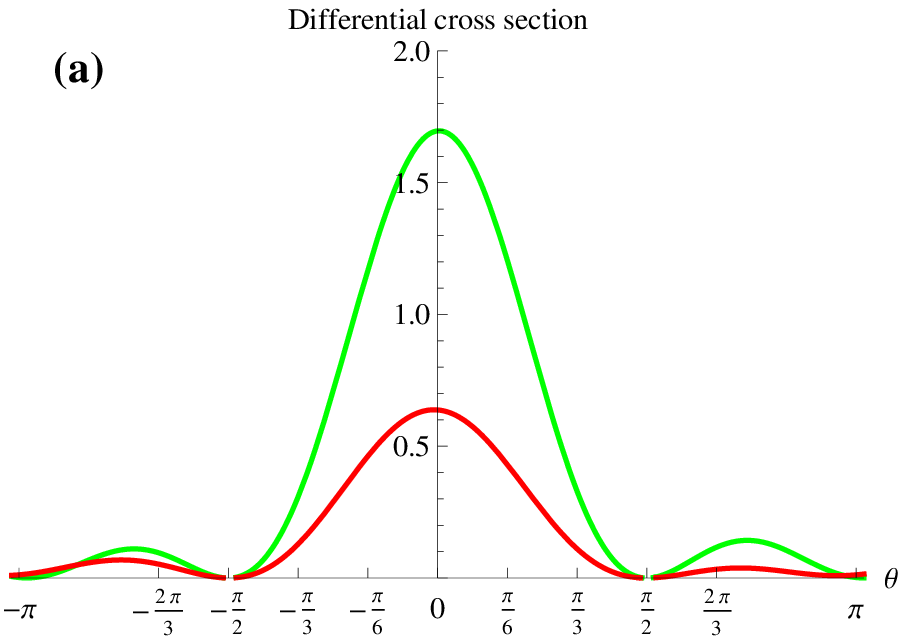}
  \ \ \includegraphics[width=8cm, height=6.5cm]{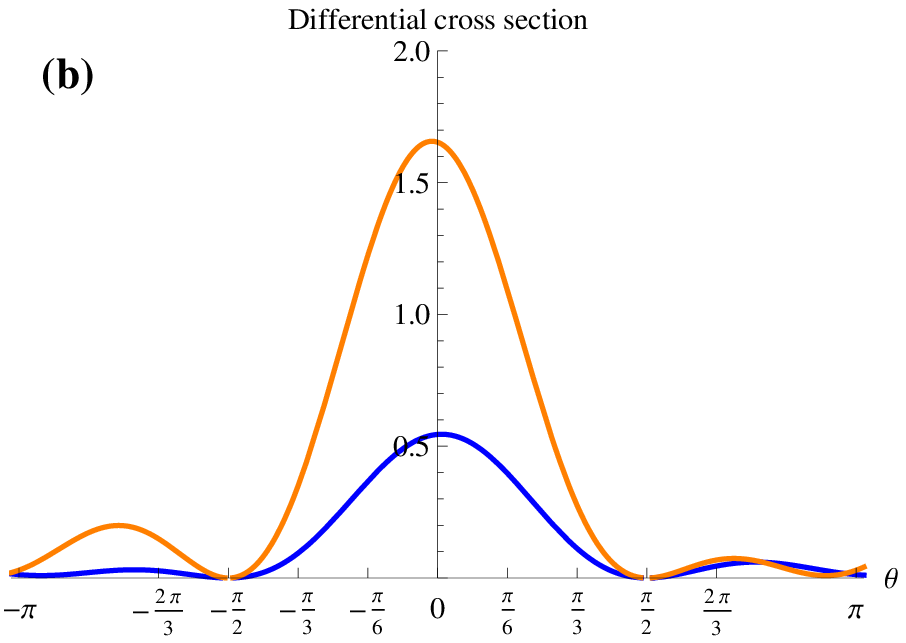}
 \caption{\sf {Angular dependence of the differential cross section,
$d\sigma(\theta)/d\theta$, in nanometers, for $R=25 nm$, $L=80
nm$, $m=0.1 meV$ and $\varepsilon=5 meV$. In (a): the green
line correspond to $U=0.2 meV$ and the pseudo scalar potential
$S=0.2 meV$. The red line correspond to $U=0.2meV$ ant %the pseudo
%scalar potential
$S=0.8 meV$. In  (b): the blue line
correspond to $U=0.2 meV$ and %the pseudo scalar potential
$S=0.8
meV$. The orange line correspond to $U=0.6meV$ and %the pseudo
%scalar potential
$S=0.8 meV$. }}\label{fig.cross11}
\end{figure}
In Figure \ref{fig.cross11} we show the angular dependence of
the differential cross section. The differential cross section
shows a narrow maximum at $\theta=0$ and has a zero minimum when
$\theta=\pi/2$. In addition the differential cross section is no
longer symmetric with respect to the sign of the incident angle. In Figure
\ref{fig.cross11}a, we fix $S$ at $0.2$ and take two different
values for $U$, $U=0.2$ correspond to the green curve and $U=0.8$
correspond to the red one. One can see that the maximum value of
the differential cross section increase when $U$ decrease to a
value that is greater than $0.2$. In Figure \ref{fig.cross11}b,
we fix $S=0.8$ and take two different values of $U$, the orange
line correspond to $U=0.2$ and the blue line correspond to
$U=0.8$. This shows that the maximum of the
differential cross section increase when the value of $U$ increase
even for values which are still smaller than $0.8$.

%%%%%%%%%%%%%%%%%%%%%%%%%%%%%%%%%%%%%%%%%%%%%%%%%%%
\section{Scattering through quantum dot}
%%%%%%%%%%%%%%%%%%%%%%%%%%%%%%%%%%%%%%%%%%%%%%%

Below we develop the scattering theory for the 2D Dirac fermions
in the presence of an axially symmetric potential using
the second potential configuration. The Dirac equation,
$H\Psi=E\Psi$, has scattering solutions that have the asymptotic
form \cite{Landau}
$\Psi(r,\theta)=\Psi_{\sf in}(r,\theta)+\Psi_{\sf out}(r,\theta)$. These
solutions have the following asymptotic forms as
$r\rightarrow\infty$
\bqr
&& \Psi_{\sf in}(r,\theta)
=\frac{1}{\sqrt{2}}e^{i\alpha r \cos\theta}  \begin{pmatrix}
       {1} \\
       {1} \\
     \end{pmatrix}\\
&& \Psi_{\sf out}(r,\theta) =\frac{e^{i\alpha r}}{\sqrt{-2i\pi
\alpha r}}\begin{pmatrix}
       {f_{+}(\theta)} \\
       {f_{-}(\theta)} \\
     \end{pmatrix}
 \eqr
where $\Psi_{\sf in}(r,\theta)$ is the incoming plane wave in the $x$-direction
and $\Psi_{\sf out}(r,\theta)$ is the scattered outgoing
spherical wave, $f_{\pm}(\theta)$ is the scattering amplitude
and $\theta$ is the scattering angle, furthermore,
$f_{-}(\theta)= e^{i\theta}f_{+}(\theta)$. The incoming particle
current density is $\alpha$. The number of particles leaving per
unit time radially in the direction $\theta$ is
$\frac{d\theta}{\pi}|f(\theta)|^2$. Thus the differential
scattering cross section is \beq \label{def-cros}
\frac{d\sigma(\theta)}{d\theta}=\frac{|f(\theta)|^2}{\pi\alpha}.
\eeq

For large $r$, the solution of the Dirac equation is given as a
function of the Hankel functions of the first and second kind,
$H^{(1,2)}_{\mu}(\alpha r)=J_{\mu}(\alpha r)\pm iY_{\mu}(\alpha
r)$, where the upper sign holds for the superscript 1 and the lower one for 2.
Their asymptotic behavior when $\alpha r \rightarrow \infty$ is given by
\bqr
&&\label{j} J_{\mu}(\alpha r)\approx \sqrt{\frac{2}{\pi \alpha
r}} \cos \left( \alpha r -\frac{(2 \mu +1)}{4}\pi \right) \\
&&
\label{jy} Y_{\mu}(\alpha r)\approx  \sqrt{\frac{2}{\pi \alpha r}}
\sin \left( \alpha r -\frac{(2 \mu +1)}{4}\pi \right) \\
&&
\label{hhh} H^{(1,2)}_{\mu}(\alpha r)\approx \sqrt{\frac{2}{\pi
\alpha r}} e^{\pm i \left( \alpha r -\frac{(2 \mu +1)}{4}\pi
\right)}.
\eqr
We note that $\Psi_{\sf out}$ for $r\rightarrow\infty$
indeed has the form of \eqref{hhh}. Using the decomposition of
the plane wave \cite{Gradshteyn}
\beq e^{i\alpha r \cos\theta}=
\sum_{j}i^{(j\mp\frac{1}{2})} e^{i(j\mp \frac{1}{2})\theta}
J_{j\mp \frac{1}{2}}(\alpha r)
\eeq
% here we have set
%$j=\frac{k}{2}$, here $j$ is a half integer
%, in the following we $m=j-\frac{1}{2}$.
gives for the incoming wave
\begin{equation}\label{psiin}
\Psi_{\sf in}=\left(%
\begin{array}{c}
  \sum_{j} i^{(j-\frac{1}{2})} J_{j-\frac{1}{2}}\left(\alpha r\right)e^{i(j-\frac{1}{2})\theta} \\
  \sum_{j} i^{(j+\frac{1}{2})} J_{j+\frac{1}{2}} \left(\alpha r\right)e^{i(j+\frac{1}{2})\theta}  \\
\end{array}%
\right).
\end{equation}
We use the defining equation
\beq f(\theta)=\frac{1}{\sqrt{2}} \sum_{j}f_j
e^{i(j \mp \frac{1}{2})\theta-i\frac{\pi}{4}}
\eeq
so that the outgoing wave function when $\alpha r \rightarrow\infty$ has the following form
\beq \label{psiout}
 \Psi_{\sf out}=\left(%
\begin{array}{c}
  \sum_{j} i^{(j-\frac{1}{2})} f_j\left[J_{j-\frac{1}{2}}\left(\alpha r\right)+iY_{j-\frac{1}{2}}\left(\alpha r\right)\right]e^{i(j-\frac{1}{2})\theta} \\
  \sum_{j} i^{(j+\frac{1}{2})}f_j\left[J_{j+\frac{1}{2}}\left(\alpha r\right)+iY_{j+\frac{1}{2}}\left(\alpha r\right)\right]e^{i(j+\frac{1}{2})\theta}  \\
\end{array}%
\right).
\eeq
We assume that the scattering defect has a finite radius $R$, so, when $r>R$ we can write the wave functions as a
superposition of terms such as
\beq \label{psij}
 \Psi_j=\left(%
\begin{array}{c}
  \left[J_{j-\frac{1}{2}}\left(\alpha r\right)+R_jY_{j-\frac{1}{2}}\left(\alpha r\right)\right]e^{i(j-\frac{1}{2})\theta} \\
  i\left[J_{j+\frac{1}{2}}\left(\alpha r\right)+R_j Y_{j+\frac{1}{2}}\left(\alpha r\right)\right]e^{i(j+\frac{1}{2})\theta}  \\
\end{array}%
\right)
\eeq
where $R_j$ is the reflection coefficients and the complex number $i$ is
introduced for further convenience. From \eqref{psiin},
\eqref{psiout} and \eqref{psij} we can write
\beq \Psi_{\sf in}+
\Psi_{\sf out}=\sum_{j} \beta_j \Psi_{j}
\eeq
and one can deduce the expression of $f_j$
 \beq f_j=\frac{R_j}{i-R_j}.
 \eeq
The expression of the scattering amplitude can then be written in compact form
 \beq f(\theta)=\sum_{j} \frac{R_j}{i-R_j} e^{i(j\mp
 \frac{1}{2})\theta-i\frac{\pi}{4}}.
 \eeq
We note that the back-scattering amplitude vanishes, $f(\pi)=0$,
which is a consequence of the pseudospin conservation for
chiral scattering \cite{Katsn09} and is related to the Klein paradox
\cite{Katsn06}.

Now we consider a circular potential barrier
in graphene and choose $V(r)=V_0\Theta(R-r)$ and
$S(r)=S_0\Theta(R-r)$ where $\Theta$ is the heaviside step
function. Using the boundary conditions, continuity of the
eigenspinors at $r=R$, we obtain
\bqr
%\begin{array}{cc}
 && {J_{j-\frac{1}{2}}\left(\alpha R\right)+R_{j}
Y_{j-\frac{1}{2}}\left(\alpha R\right)=T_jJ_{j-\frac{1}{2}}\left(\alpha' R\right)}\\
&& {J_{j+\frac{1}{2}}\left(\alpha R\right)+R_j
Y_{j+\frac{1}{2}}\left(\alpha
R\right)=T_jJ_{j+\frac{1}{2}}\left(\alpha' R\right)}
\eqr
where $\alpha=\sqrt{\left(\varepsilon-V_0 \right)^2-\left(
\Delta+S_0 \right)^2}$ and $\alpha'=\sqrt{\varepsilon^2-
\Delta^2}$. The reflection coefficient is then given by
\beq
R_j=-\frac{J_{j-\frac{1}{2}}\left(\alpha
R\right)J_{j+\frac{1}{2}}\left(\alpha'
R\right)-J_{j+\frac{1}{2}}\left(\alpha
R\right)J_{j-\frac{1}{2}}\left(\alpha' R\right)}
{Y_{j-\frac{1}{2}}\left(\alpha
R\right)J_{j+\frac{1}{2}}\left(\alpha'
R\right)-Y_{j+\frac{1}{2}}\left(\alpha
R\right)J_{j-\frac{1}{2}}\left(\alpha' R\right)}.
\eeq
Similarly for the transmission coefficient %is given by
\beq
T_j=\frac{Y_{j+\frac{1}{2}}\left(\alpha
R\right)J_{j-\frac{1}{2}}\left(\alpha
R\right)-Y_{j-\frac{1}{2}}\left(\alpha
R\right)J_{j+\frac{1}{2}}\left(\alpha R\right)}
{Y_{j+\frac{1}{2}}\left(\alpha
R\right)J_{j-\frac{1}{2}}\left(\alpha'
R\right)-Y_{j-\frac{1}{2}}\left(\alpha
R\right)J_{j+\frac{1}{2}}\left(\alpha' R\right)}.
\eeq
We have the relations $J_{-n}(x)=(-1)^n J_n(x)$ and $Y_{-n}(x)=(-1)^n Y_n(x)$
\cite{Gradshteyn}. If we take $n=j-\frac{1}{2}$, $n$ is an integer,
we get $R_{n}=R_{-n-1}$, so the back-scattering amplitude
vanishes, and $T_{n}=T_{-n-1}$, that is $n \leftrightarrow -n-1$ which
means $f_{n}=f_{-n-1}$. Thus,  \eqref{def-cros} can be
rewritten in a compact form
\beq
\frac{d\sigma(\theta)}{d\theta}=\frac{1}{\pi \alpha} \left|
\sum_{n=0}^{\infty} f_{n} \cos{(n+\frac{1}{2})\theta} \right|^{2}.
\eeq

The angular dependence of the cross section is shown in Figure
\ref{fig.cross6}, where we have the cross section as a function of the incident angle $\theta$
which shows a narrow maximum at $\theta = 0$.\\

\begin{figure}[h]
\centering
  \includegraphics[width=8cm, height=6.5cm]{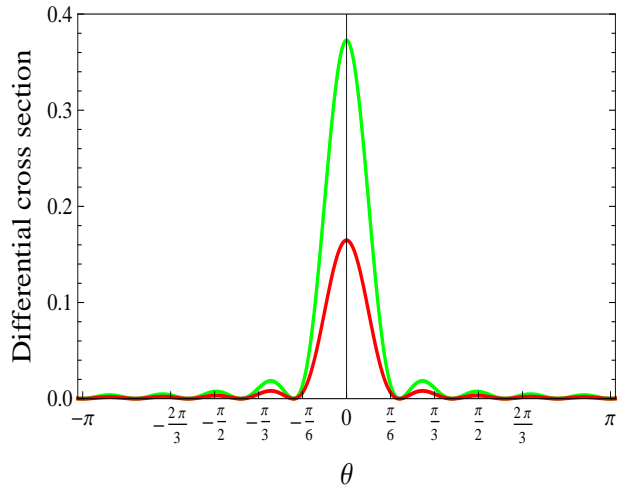}
 \caption{\sf {Angular dependence of the differential cross section,
$d\sigma(\theta)/d\theta$, in nanometers, for the radius $R=15nm$
with the vector potential $V_0=40meV$. The green line correspond
to $\varepsilon=20meV$, the pseudo scalar potential $S_0=8meV$ and
with the mass term $m=7meV$. The red line correspond to
$\varepsilon=15meV$, %the pseudo scalar potential
$S_0=5meV$ and
with the mass term $m=2.5meV$}}\label{fig.cross6}
\end{figure}
In Figure \eqref{fig.cross6} we plot the dependence of the
differential cross section on the incident angle $\theta$
choosing different values of the parameters. We show that the
curve is symmetric about the axis $\theta=0$. In addition, it is
clearly seen that the differential cross section exhibits a narrow
maximum around $\theta=0$ and vanishes when $\theta$ goes to $\pm \pi$.
The differential cross section shows resonances associated with the QD quasi-bound states \cite{Katsn09}.

%We note that for the cross section, we have the regimes that
%depending on whether $\alpha^{'}R\gg 1 $ or $\alpha^{'}R \ll 1 $

%In the first case we use the asymptotic expressions for the Bessel
%functions at $\alpha^{'}r\rightarrow \infty$, then the expression
%of $R_j$ is simplifies to \beq R_j=\tan(\alpha^{'}-\alpha) \eeq in
%this approximation the cross section can by writing as \beq
%\frac{\sigma(\theta)}{d \theta}\approx  \sum_{j,j'}\frac{2}{\pi
%\alpha}|\sin(\alpha^{'}-\alpha)|^2e^{i(j-j')\theta} \eeq then the
%transport cross section become \beq \sigma_{tr}=
%\int_{-\pi}^{\pi}d\theta
%[1-\cos(\theta)]\frac{d\sigma(\theta)}{d\theta}\propto
%\frac{|\sin(\alpha^{'}-\alpha)|^2}{\alpha} \eeq
% in this
%case we note that the cross section has a maximum when $\theta=0$

% In the second cases the Born approximation can be
%used
% We expand the Bessel
%functions for $j=\frac{1}{2}$ that correspond with
%$j=-\frac{1}{2}$ to the largest values of $R_j$ when
%$R\rightarrow0$, and we use the fact that $\lim\limits_{x \to
%0}J_0(x)\approx 1$, $\lim\limits_{x \to 0}J_1(x)\approx
%\frac{x}{2}$, $\lim\limits_{x \to 0}Y_0(x)\approx
%\frac{2}{\pi}\left[\log(\frac{x}{2}+\gamma) \right]$ and
%$\lim\limits_{x \to 0}Y_1(x)\approx -\frac{2}{\pi x}$ we obtain
%\beq \sigma\left(\theta\right)\thickapprox .... \eeq

%\frac{2\left(\alpha' R\right)^2}{\pi k\log^2\left(\alpha
%R\right)}\left[1-\cos\left(\theta\right)\right]

%%%%%%%%%%%%%%%%%%%%%%%%%%%%%%%%%%%%%%%%%%%%%%%%%%%
\section{Conclusion}
%%%%%%%%%%%%%%%%%%%%%%%%%%%%%%%%%%%%%%%%%%%%%%%

We  studied the Dirac equation in $(2+1)$-dimensions where we included all types of
potential couplings: vector, pseudo-scalar and scalar. We used
the method of separation of variables in polar coordinates to obtain the general spinor
eigenfunctions and associated energy spectra for two special potential
configurations. We have studied the bound states of graphene QD
which can be created electrostatically in presence of a constant
mass term. We discussed the character of the electronic eigenstates
occurring in various regions of energy and potential parameters.

We used the solution of the Dirac equation for two potential
configurations and studied the dependence of the electronic bound
states on the strength of the electrostatic potential and the radius of the QD. For
both potential configurations we showed that when we increase the radius of the QD
or the strength of the confining potential more bound states can be
accommodated in the QD, in agreement with usual quantum mechanical results.

We developed the scattering theory for the 2D Dirac
fermions in the presence of the axially symmetric potential and
computed the differential cross section. This differential cross section is
associated with a quantum dot embedded in an environment. We showed that it
is no longer symmetric with respect to the sign of the incident angle, it
has a maximum around $\theta=0$ and exhibits resonances associated with quantum dot
quasi-bound states. We also found that the variation of the
potential strength changes the maximum of the differential cross section.

%%%%%%%%%%%%%%%%%%%%%%%%%%%%%%%%%%%%%%%%%%%%%%%%%%%
%\section{Appendix}
%%%%%%%%%%%%%%%%%%%%%%%%%%%%%%%%%%%%%%%%%%%%%%%

%%%%%%%%%%%%%%%%%%%%%%%%%%%%%%%
\section*{Acknowledgments}
%%%%%%%%%%%%%%%%%%%%%%%%%%%%%

The generous support provided by the Saudi Center for Theoretical
Physics (SCTP) is highly appreciated by all authors. AJ and HB
acknowledges partial support by King Fahd University of Petroleum and minerals
under the theoretical physics research group project RG1306-1 and RG1306-2.

\end{document}